\documentclass[12pt]{article}

\usepackage{amsmath,amssymb,amsfonts,amsthm,bbm}
\usepackage{mathtools}   

\usepackage{enumitem}   
\usepackage{multirow}
\usepackage{authblk} 
\usepackage{threeparttable}
\usepackage{caption}

\newtheorem{theorem}{Theorem}

\usepackage[titletoc,title]{appendix}  

\usepackage{kpfonts}
\usepackage[T1]{fontenc}

\usepackage[table,xcdraw,dvipsnames]{xcolor}   

\usepackage{hyperref}
\newcommand\myshade{85}
\colorlet{mylinkcolor}{YellowOrange}
\colorlet{mycitecolor}{Aquamarine}
\colorlet{myurlcolor}{violet}

\hypersetup{
linkcolor  = mylinkcolor!\myshade!black,
citecolor  = mycitecolor!\myshade!black,
urlcolor   = myurlcolor!\myshade!black,
colorlinks = true,
}

\usepackage{caption}
\usepackage{subcaption}
\DeclarePairedDelimiter{\abs}{\lvert}{\rvert}

\usepackage[linesnumbered,ruled,vlined]{algorithm2e}

\SetCommentSty{mycommfont}

\SetKwInput{KwInput}{Input}                
\SetKwInput{KwOutput}{Output}              

\usepackage{listings}             
\renewcommand{\hat}{\widehat}
\renewcommand{\tilde}{\widetilde}

\newcommand{\bfm}[1]{\ensuremath{\boldsymbol{#1}}} 

   \def\bA{\bfm A}  
   \def\bB{\bfm B}  
   \def\bC{\bfm C}  
     
\def\be{\bfm e}   \def\bE{\bfm E}  \def\EE{\mathbb{E}}
\def\bff{\bfm f}  \def\bF{\bfm F}  
   \def\bG{\bfm G}  
   \def\bH{\bfm H}  
   \def\bI{\bfm I}

   \def\bM{\bfm M}

   \def\bQ{\bfm Q}  
   \def\bR{\bfm R}  \def\RR{\mathbb{R}}
   \def\bS{\bfm S}  
     
\def\bu{\bfm u}   \def\bU{\bfm U}  
   \def\bV{\bfm V}  
     
   \def\bX{\bfm X}  
   \def\bY{\bfm Y}

\def\calD{{\cal  D}}

\def\calN{{\cal  N}} 
\def\calO{{\cal  O}}

\def\calU{{\cal  U}}

\newcommand{\bfsym}[1]{\ensuremath{\boldsymbol{#1}}}

\def\balpha{\bfsym \alpha}

             \def\bGamma{\bfsym \Gamma}

\def\bepsilon{\bfsym \varepsilon}
             \def\bSigma{\bfsym \Sigma}
        
          \def\bOmega {\bfsym {\Omega}}

          \def\bXi{\bfsym {\Xi}}

          \def\bPhi{\bfsym {\Phi}}
\def\bPsi{\bfsym{\Psi}}

\providecommand{\abs}[1]{\left\lvert#1\right\rvert}
\providecommand{\norm}[1]{\left\lVert#1\right\rVert}

\providecommand{\paran}[1]{\left( #1 \right)}

\newtheorem{assumption}{Assumption}

\newtheorem{remarkx}{Remark}

\newcommand{\E}[1]{{\mathbb{E}} \left[ #1 \right]}    
\newcommand{\Var}[1]{{\rm Var} \left[ #1 \right]}
\newcommand{\Cov}[1]{{\rm Cov}  \left[ #1 \right]}

\newcommand{\vect}[1]{{\textsc{vec}} \left( #1 \right)}

\newcommand{\Tr}[1]{{\rm Tr} \left( #1 \right) }
\newcommand{\Op}[1]{{\calO_p} \left( #1 \right) }

\newcommand{\bigO}[1]{{\calO} \left( #1 \right) }

\newcounter{question}

\numberwithin{equation}{section}  

\newcounter{CondCounter}

\usepackage[table,xcdraw,dvipsnames]{xcolor}   
\definecolor{royalpurple}{rgb}{0.47, 0.32, 0.66}

\renewcommand{\bar}{\overline}

\usepackage{setspace}
\usepackage{etoolbox}
\BeforeBeginEnvironment{equation*}{\begin{singlespace}\vspace*{-\baselineskip}}
\AfterEndEnvironment{equation*}{\end{singlespace}\noindent\ignorespaces}
\BeforeBeginEnvironment{equation}{\begin{singlespace}\vspace*{-\baselineskip}}
\AfterEndEnvironment{equation}{\end{singlespace}\noindent\ignorespaces}
\BeforeBeginEnvironment{align}{\begin{singlespace}\vspace*{-\baselineskip}}
\AfterEndEnvironment{align}{\end{singlespace}\noindent\ignorespaces}
\BeforeBeginEnvironment{align*}{\begin{singlespace}\vspace*{-\baselineskip}}
\AfterEndEnvironment{align*}{\end{singlespace}\noindent\ignorespaces}
\BeforeBeginEnvironment{eqnarray}{\begin{singlespace}\vspace*{-\baselineskip}}
\AfterEndEnvironment{eqnarray}{\end{singlespace}\noindent\ignorespaces}
\BeforeBeginEnvironment{eqnarray*}{\begin{singlespace}\vspace*{-\baselineskip}}
\AfterEndEnvironment{eqnarray*}{\end{singlespace}\noindent\ignorespaces}

\usepackage[framemethod=TikZ]{mdframed} 

\usepackage{fancybox}

\usepackage{setspace}

\def\spacingset#1{\renewcommand{\baselinestretch}%
{#1}\small\normalsize} \spacingset{1}

\usepackage[authoryear]{natbib}  
\usepackage{comment}
\usepackage{silence}
\usepackage{graphicx}
\graphicspath{{figs/}}

\newcommand\lnorm{\left\lVert}
\newcommand\rnorm{\right\lVert}
\begin{document}
%
%

\def\TITLE{Time-Varying Matrix Factor Models}

\newcommand{\blind}{0}

\if0\blind
{
\title{\bf \TITLE\footnote{We thank Zeyu Wang and Yucheng Wong for their excellent research assistance. We also thank John Chao, Giovanni Motta, Mahrad Sharifvaghefi, Mark Watson, and seminar participants at Boston College, University of Maryland, Purdue University, Vanderbilt University, University of California, Riverside, the 2023 NBER-NSF Time Series Conference and the 15th Greater New York Metropolitan Area Econometrics Colloquium for their useful comments and discussions. Any remaining errors are
solely ours. R. Chen's research is supported
in part by National Science Foundation
grants DMS-2027855, DMS-2052949 amd DMS-2319260.}}
\author[1]{Bin Chen\thanks{binchen@rochester.edu}}
\author[2]{Elynn Y. Chen \thanks{elynn.chen@stern.nyu.edu}} 
\author[3]{Stevenson Bolivar\thanks{sb2230@stat.rutgers.edu}} 
\author[3]{Rong Chen\thanks{rongchen@stat.rutgers.edu}}
\affil[1]{Department of Economics, University of Rochester}
\affil[2]{Stern School of Business, New York University} 
\affil[3]{Department of Statistics, Rutgers University}
\date{\today}
\maketitle
} \fi

\if1\blind
{
\bigskip
\bigskip
\bigskip
\title{\bf \TITLE}
\author[1]{}
\date{\vspace{10em}}
\maketitle
\medskip
} \fi

%
%
\begin{abstract}
Matrix-variate data of high dimensions are frequently observed in finance and economics, spanning extended time periods, such as the long-term data on international trade flows among numerous countries. To address potential structural shifts and explore the matrix structure's informational context, we propose a time-varying matrix factor model. This model accommodates changing factor loadings over time, revealing the underlying dynamic structure through nonparametric principal component analysis and facilitating dimension reduction. We establish the consistency and asymptotic normality of our estimators under general conditions that allow for weak correlations across time, rows, or columns of the noise. A novel approach is introduced to overcome rotational ambiguity in the estimators, enhancing the clarity and interpretability of the estimated loading matrices.  Our simulation study highlights the merits of the proposed estimators and the effective of the smoothing operation. In an application to international trade flow, we investigate the trading hubs, centrality, patterns, and trends in the trading network.
\bigskip


\bigskip

\bigskip
\end{abstract}

%
%

\spacingset{1.8} 

\section{Introduction}

Matrix-variate data are increasingly prevalent across various fields, such as economics, finance, and social networks. Two significant types of such data are particularly noteworthy: Dynamic transportation networks are represented by matrices of monthly import-export volumes between countries, highlighting the complex, weighted, and directional relationships. Dynamic panel data, vital for investor portfolio decisions, are matrices capturing various firm attributes (e.g., stock prices, book-to-market ratio, dividend-to-price ratio) at different times, capturing firm performance and market position.

The field of statistical analysis for matrix-variate data is still emerging, with matrix factor models being pivotal for dimension reduction and uncovering inherent structures. \cite{wang2019factor} introduced a matrix factor model for time series, featuring constant matrix factor loadings and time-varying latent factors, estimated through eigenanalysis of auto-cross-covariance matrices. This model has been expanded into a constrained version by \cite{chen2019constrained} and a threshold variant by \cite{liu2019threshold}. Contrasting these approaches, \cite{chen2021statistical} developed a novel estimation method by analyzing an aggregate of the sample mean matrix and contemporary covariance matrices, proving to be advantageous in several contexts.

While matrix factor models have been proven effective, a common assumption in the literature is that factor loadings remain constant over time -- an assumption not always realistic in practice. Changes in policies, preferences, technology, and environmental factors can influence economic variables' relationships, challenging the stability of factor loadings. Empirical evidence, such as that presented by \cite{chen2019factor}, indicates that ignoring these potential shifts in factor loadings can result in inconsistent estimates and unreliable conclusions.

Recently, the concept of smooth structural changes in economic variables has garnered attention, leading to the development of time-varying time series and panel data models. These models, particularly nonparametric time-varying parameter models introduced by \cite{robinson1989nonparametric} and further explored by \cite{cai2007trending}, \cite{chen2012testing}, \cite{zhang2012inference}, \cite{inoue2017}, and \cite{chen2022}, offer a flexible approach to modeling economic variables. These models stand out for their minimal assumptions on the time-varying parameters' functional forms, requiring only that they change smoothly over time. Inspired by the versatility and effectiveness of this approach, we employ this framework to propose time-varying matrix factor models for analyzing matrix time series.

In this paper, we investigate a matrix factor model with loading matrices defined as unknown, smooth, time-varying functions. We aim to estimate these functions and the latent factors using local Principal Component Analysis (PCA), developing a theory of inference that includes consistency, convergence rates, and limiting distributions. This approach accommodates time-varying loadings and weak correlations across time, rows, or columns of the noise matrices. We generalize the eigenvalue ratio-based estimator for latent dimensions \citep{ahn2013} to our time-varying context and establish the consistency of the new method. 

Traditional methods reduce matrix observations into vectors and apply local PCA to high-dimensional factor models \citep{su2017time}, losing essential matrix structure and leading to less effective analyses. Our matrix-based estimator is compared against vectorized approaches through simulations to demonstrate its superior ability to preserve and leverage the matrix structure. Additionally, we tackle the rotation ambiguity inherent in matrix factor models by introducing a smoothing process for the time-varying loading spaces, facilitating clearer interpretation.

This paper is organized as follows: Section \ref{sec:model} introduces matrix factor models with time-varying loadings. Section \ref{sec:estimation} describes our local PCA estimation approach and a generalized eigenvalue ratio-based estimator for latent dimensions. Section \ref{sec-theory} discusses the theoretical foundations, including consistency and asymptotic normality. Section \ref{sec:smooth} details the smoothing process for time-varying functions. Section \ref{sec-simul} assesses our methods through simulation, while Section \ref{sec-appl-it} applies them to international trade data. All mathematical proofs and additional results are provided in the Appendix.

\section{Matrix Factor Models with Time-Varying Loadings} \label{sec:model}

Let $\bY_t$, $t\in[T]$, be a matrix-valued time series, where each $\bY_t=[Y_{t,ij}]$ is a matrix of size $p \times q$.
We consider the following matrix factor models with time-varying factor loadings, 

\begin{equation} \label{eqn:mfm}
	\bY_t = \bR_t \bF_t \bC^\top_t + \bE_t,  \quad t\in[T],
\end{equation}
where $\bF_t$ is the $k \times r$ common factor matrix, $\bR_t$ ($\bC_t$) is a $p \times k$ ($q \times r$) time-varying row (column) loading matrix and the sequence of matrices $\bE_t$ is the noise matrix. 
The noise term $\bE_t$ is assumed to be uncorrelated with $\bF_t$, yet is allowed to be weakly correlated across rows, columns, and times. 
It generalizes the matrix factor model: 
\begin{equation}\label{eq:matrix_factor}
	\bY_t=\bR\bF_t\bC^\top+\bE_t,
\end{equation}
by allowing for structural changes in factor loadings $\bR$ and $\bC$. To cover a wide range of potential smooth temporal variation, we follow the literature on smooth time-varying parameter models \citep{robinson1989nonparametric,cai2007trending,su2017time}) and model $\bR_{t,i.}$ ($\bC_{t,j.}$) as a non-stochastic function of $t/T$, that is,
\[
\bR_{t,i.} = \bR_{i \cdot}(t/T)  \mbox{ \ \ and \ \ }
\bC_{t,j.} =  \bC_{j \cdot}(t/T), 
\]
where $\bR_{t,i.}(\cdot)$ ($\bC_{t,j.}(\cdot)$) is an unknown smooth function of $t/T$ on $[0, 1]$ for each $i$ ($j$). The specification that loading matrices $\bR_{t,i.}(\cdot)$ and $\bC_{t,j.}(\cdot)$ are some functions of ratio $t/T$ rather than time $t$ only is a common scaling scheme in the literature for the ease of theoretical analysis \citep{phillips1990,robinson1991nonparametric,cai2007trending}. 
We note that model \eqref{eqn:mfm} {is subject to} scale and rotational ambiguity in its individual components $\bR_t$, $\bF_t$ and $\bC_t$. {Our objective is} to estimate sensible representations of $\bR_t$ and $\bC_t$ among their equivalent classes and the corresponding {$\bF_t$}, as the signal {$\bS_t=\bR_t\bF_t\bC_t^\top$} is uniquely defined. 
\begin{figure}[ht!]
	\begin{center}
		\caption{Time series plots of the value of goods traded among 24 countries.}
		\begin{threeparttable}
			\includegraphics[width=\linewidth,height=\textheight,keepaspectratio=true]{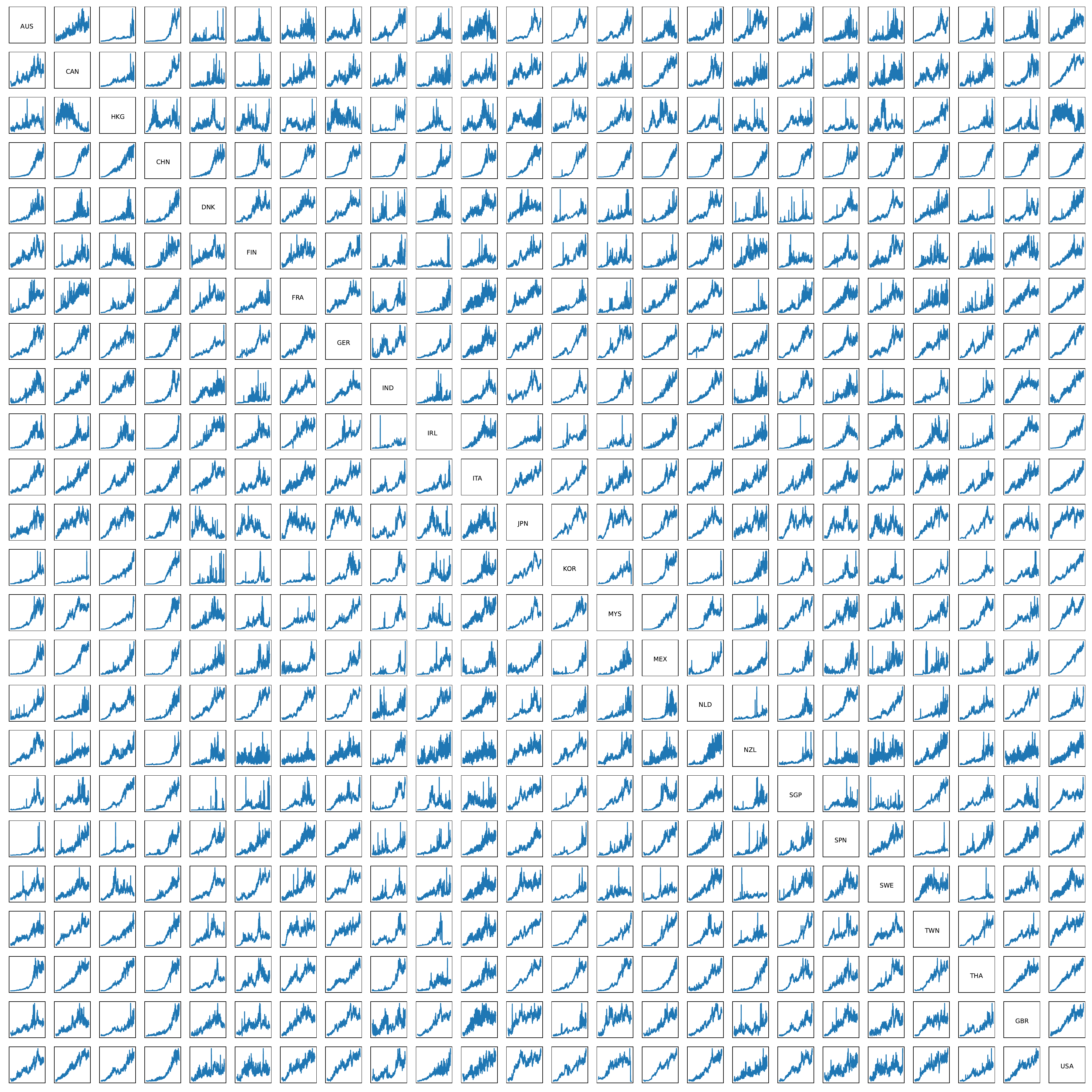}
			\begin{tablenotes}
				\small
				\item Notes:  (1) sample period: January 1982- December 2018. (2) The plots only show the patterns of the time series while the magnitudes are not comparable between plots because the ranges of the y-axis are different.
			\end{tablenotes}
		\end{threeparttable}
		\label{fig:time_series}
	\end{center}
\end{figure}

\begin{figure}[ht!]
	\caption{(left) Estimated transport network, June, 2016, and (right) the evolution of one row vector of $\bR_t$ with a 5-year rolling window.}
	\centerline{\includegraphics[width=0.4\linewidth,height=2.6in]{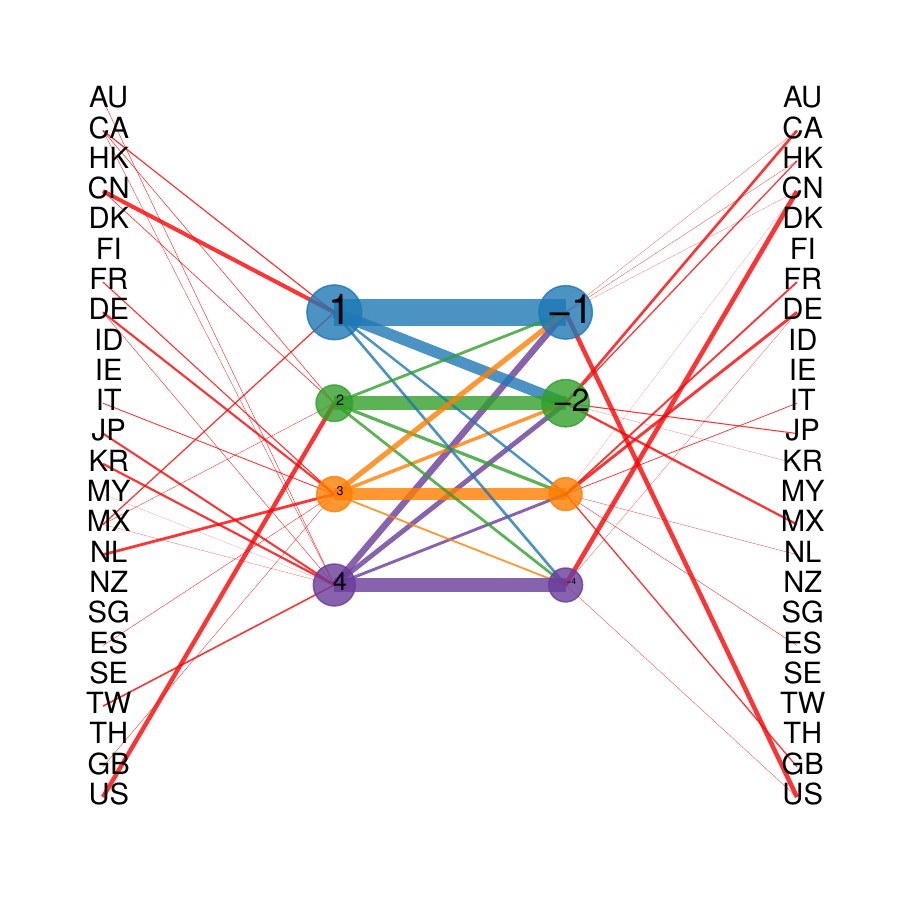}
		\includegraphics[width=0.5\linewidth,height=2.3in]{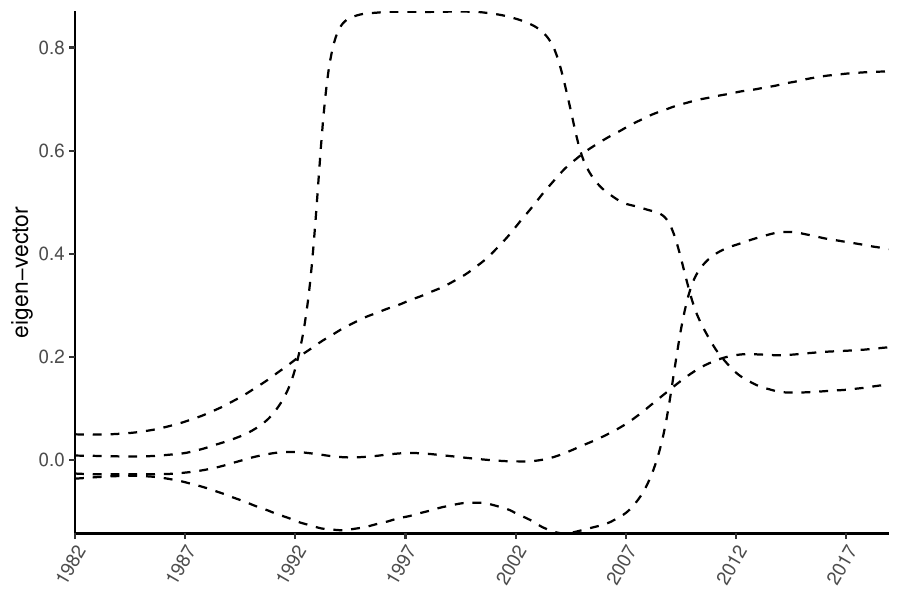}
	}
	\label{fig:transport_network}
\end{figure}
As a motivating example, we introduce the international trade flow time series, with a detailed analysis in Section~\ref{sec-appl-it}. The data, $\mathbf{Y}_{t}$, is represented as a square matrix for each time $t$, where each element, $y_{t,ij}$, indicates the trade volume from country $i$ (exporter) to country $j$ (importer). The time series, spanning from January 1982 to December 2018, depicted in Figure 1, showcases the non-stationary nature of the data, highlighting a globalization trend. \cite{chen2019modeling} interpreted the matrix factor model for transport data, viewing the factor process $\bF_t$ as representing trade volumes among {\bf (latent) trading hubs}, analogous to trade volumes among groups of countries. This conceptualization suggests that a country's exports pass through these trading hubs before reaching the importing country, with $\bF_t$ detailing hub-to-hub trade activities. The loading matrices $\bR_t$ and $\bC_t$ indicate each country's involvement with the hubs, with elements like $\bR_{t,11}$ showing country 1's interaction with export hub 1. The cumulative export volume of a country, such as country 1, is calculated as $\sum_{k=1}^r\bR_{t,1k}\bF_{t,k1}$. A visual representation for June 2016 (left figure in Figure~\ref{fig:transport_network}) demonstrates this model, where hub sizes reflect trade volume $\bF_t$ and line widths show participation levels $\bR_t$ and $\bC_t$, with minor values (less than 0.15) excluded for clarity. This visualization reveals, for example, significant trade between China’s primary export hub and the main import hubs (Hub 1 and 2) used by the US, Canada, Mexico, Japan, Korea, and Taiwan in that month. The right figure in Figure~\ref{fig:transport_network} illustrates changes over time in China's export loading vectors, highlighting structural shifts in trading patterns based on a five-year moving analysis. 

\section{Estimation}  \label{sec:estimation}

The time-varying nature of \eqref{eqn:mfm} presents a significant challenge in estimating $\bR_{t}$, $\bC_{t}$, and $\bF_t$, as there is effectively only one sample point available at each time $t$. To see this challenge, we first analyze the problem on the population level. Under the time-varying matrix factor model \eqref{eqn:mfm},  the second {\em uncentered} moment writes
\begin{align}
	\E {\bY_t\bY_t^\top} & = \bR_t \E{\bF_t \bC_t^\top\bC_t\bF_t^\top}\bR_t^\top + \E {\bE_t \bE_t^\top}, \label{eqn:mfm-2nd-moment-R} \\
	\E {\bY_t^\top\bY_t} & = \bC_t \E{\bF_t^\top \bR_t^\top\bR_t\bF_t}\bC_t^\top + \E {\bE_t^\top \bE_t}, \label{eqn:mfm-2nd-moment-C}
\end{align}
which aggregate the information of unknown parameters $\bR_t$ and $\bC_t$ from both the first moment $\E{\bY_t} = \bR_t \E{\bF_t} \bC_t^\top$ and the second {\em centered} moment $\E{\paran{\bY_t-\E {\bY_t}} \paran{\bY_t-\E{\bY_t}}^\top }$. 

Noticing that the first term on the right-hand-side of \eqref{eqn:mfm-2nd-moment-R} is of rank $k$ under some mild conditions and ignoring the second term (as justified by the pervasive assumption below), it is easy to see $\bR_t$ is the same as the top $k$ eigenvectors of the second moment, up to an affine transformation. Similar observation holds for $\bC_t$ with \eqref{eqn:mfm-2nd-moment-C}. 

This motivates the employment of the spectral method based on the sample estimators of $\E{\bY_t\bY_t^\top}$ and $\E{\bY_t^\top\bY_t}$ to estimate $\bR_t$ and $\bC_t$. 
As model \eqref{eqn:mfm} is time varying, we use nonparametric kernel smoothing to estimate $\E{\bY_t\bY_t^\top}$ and $\E{\bY_t^\top\bY_t}$, based on the smoothness assumption of the loading functions $\bR_{i\cdot}(t/T)$ and 
$\bC_{i\cdot}(t/T)$ and the stationary assumptions of the factor 
process $\bF_t$, which leads to smoothness of $\E{\bY_t\bY_t^\top}$ and $\E{\bY_t^\top\bY_t}$ in $t/T$.
Specifically, we define sample scatter matrix $\hat{\bM}_{R,t}$ and $\hat{\bM}_{C,t}$ as 
\begin{equation} \label{eqn:mhat}
	\hat{\bM}_{R,t}  = \frac{1}{pqT} \sum_{s=1}^{T} K_{h,ts}\bY_s\bY_s^\top
	\\
	\mbox{ \ \ and \ \ }
	\hat{\bM}_{C,t}  = \frac{1}{pqT} \sum_{s=1}^{T} K_{h,ts}\bY_s^\top\bY_s
\end{equation}
where $K_{h,st}$ is a boundary-modified kernel function: 
\begin{equation}
	\label{eqn:boundary_kernel}
	K_{h,st}=\left\{
	\begin{array}{ll}
		h^{-1} k(\frac{s-t}{Th})/\int_{-t/\left(Th\right)}^{1} k\left(u\right) \,du & if t\in\left[1,\lfloor Th\rfloor\right],\\
		h^{-1} k(\frac{s-t}{Th}) & if t\in\left[\lfloor Th\rfloor,T-\lfloor Th\rfloor\right],\\
		h^{-1} k(\frac{s-t}{Th})/\int_{-1}^{\left(1-t/T\right)/h} k\left(u\right) \,du & otherwise.
	\end{array}
	\right.\end{equation}
The kernel $k(\cdot): [-1,1] \to \RR^+$ is a predetermined symmetric probability density function, with $h = h(p, q, T)$ as the bandwidth parameter for the row (column) loading matrix $\bR_t$ ($\bC_t$), and $\lfloor Th\rfloor$ representing the integer part of $Th$. Common examples of $k(\cdot)$ are the uniform, Epanechnikov, and quartic kernels \citep{hastie2009elements}.

To estimate $\bR_{t}$, and $\bC_{t}$, we use $\sqrt{p}$ ($\sqrt{q}$) multiplied by the eigenvectors  corresponding to the $k$ ($r$) largest eigenvalues of $\hat{\bM}_{R,t}$ ($\hat{\bM}_{C,t}$). 
Due to rotational ambiguity, the solution $\hat{\bR}_t$ ($\hat{\bC}_t$) is one of the representatives of the column space of $\bR_t$ ($\bC_t$). They do not necessarily constitute a smooth function $\bR_t=\bR(t/T)$ as $t$ changes. In Section~\ref{sec:smooth} we present a
smoothing procedure to obtain a smooth 
representation. Furthermore, the realized latent factor $\bF_t$ and 
the signal part $\bS_t$ can be obtained by 
\begin{equation*}
	\hat \bF_t = \frac{1}{p q} \hat {\bR}_t^\top \bY_t \hat{\bC}_t 
	\mbox{\ \ and \ \ }
	\hat \bS_t = \frac{1}{p q} \hat{\bR}_t \hat{\bR}_t^\top \bY_t \hat{\bC}_t \hat{\bC}_t^\top.
\end{equation*}

The above local-PCA-based method assumes that the latent dimensions $k\times r$ are known. However, in practice, we need to estimate $k$ and $r$ as well. We extend the eigenvalue ratio-based estimator, proposed by \cite{ahn2013}. Let $\hat{\lambda}_{1,t} \geq \hat{\lambda}_{2,t} \geq \cdots \geq \hat{\lambda}_{k,t} \geq 0$ be the ordered eigenvalues of $\hat{\bM}_{R,t}$. The generalized eigenvalue ratio-based estimator for $k$ can be defined as 
\begin{equation} \label{eqn:eigen_ratio}
	\hat{k} = \underset{1 \le j \le k_{max} }{\text{argmax}} \frac{1}{T} \sum_{t=1}^{T}\frac{\hat{\lambda}_{j,t}}{\hat{\lambda}_{j+1,t}}, 
\end{equation}
and $k_{max}$ is a selected upper bound. We could choose $k_{max}=\lfloor p/2 \rfloor $ or $k_{max}=\lfloor p/3 \rfloor $. The estimator $\hat{r}$ can be defined analogously with respect to $\hat{\bM}_{C,t}$.

\section{Theory} \label{sec-theory}

Let $\bR_t^{\left(1\right)}$ $\left(\bC_t^{\left(1\right)}\right)$ and $\bR_t^{\left(2\right)}$ $\left(\bC_t^{\left(2\right)}\right)$ denote the first and second order derivative of $\bR_t$ $\left(\bC_t\right)$ respectively, and $\bR_t^{\left(0\right)}=\bR_t$ and $\bC_t^{\left(0\right)}=\bC_t$. To derive the asymptotic properties of our estimators, we impose the following regularity conditions.

\begin{assumption}[\textbf{$\alpha$-mixing}] \label{assume:alphamixing}
The vector-variate process 
$\{\vect{\bF_t}^\top,\vect{\bE_t}^\top\}^\top$ 
is a $\alpha$-mixing process with mixing coefficients 
$\alpha(h)=\underset{\tau}{\sup} \underset{A \in \mathcal{F}_{-\infty}^{\tau}, B \in \mathcal{F}_{\tau+h}^{\infty}}{\sup} \left| P(A \cap B) - P(A)P(B) \right|$, where $\mathcal{F}_{\tau}^s$ is the $\sigma$-field generated by $\{\vect{\bF_t}^\top,\vect{\bE_t}^\top\}^\top: {\tau} \le t \le s \}$, and the mixing coefficients satisfy the condition that
$\sum_{h=1}^{\infty} \alpha(h)^{1-2/\gamma} < \infty$, 
for some $\gamma>2$.
\end{assumption}

\begin{assumption}[\textbf{Factor and noise matrices}] \label{assume:factor-noise-matrix}
There is a positive constant $C < \infty$ such that 
\begin{enumerate}[label=(\alph*)]
\item Factor matrix $\bF_t$ is of fixed dimension $k \times r$, and $\max_t\EE\norm{\bF_t}^4 \le C$.

\item For all $i \in [p]$, $j \in [q]$ and $t \in [T]$, $\E{e_{t, ij}} = 0$ and $\EE \lvert e_{t, ij} \rvert^8 \le C$.

\item Factor and noise are uncorrelated, that is, $\E{e_{t,ij} f_{s,lh}} = 0$ for any $t, s \in [T]$, $i\in[p]$, $j\in [q]$, $l\in[k]$, and $h\in[r]$.
\end{enumerate}
\end{assumption}

\begin{assumption}[\textbf{Loading matrix}] \label{assume:loading_matrix}
There exists a positive constant $C < \infty$ such that 
\begin{enumerate}[label=(\alph*)]
\item For each row of $\bR_t$, $\max_t\left\Vert{\bR_{t,i \cdot}} \right\Vert\le C$, and, as $p,q \rightarrow \infty$, we have $\max_t\left\Vert{p^{-1} {\bR_t}^\top \bR_t - \bOmega_{R,t}} \right\Vert \longrightarrow 0$ for some $k \times k$ positive definite matrix $\bOmega_{R,t}$.
\item For each row of  $\bC_t$,  $\max_t\left\Vert{\bC_{t,i \cdot}} \right\Vert\le C$, and, as $p,q \rightarrow \infty$, $\max_t
\left\Vert{q^{-1} {\bC_t}^\top \bC_t - \bOmega_{C,t}}\right\Vert \longrightarrow 0$ for some $r \times r$ positive definite matrix $\bOmega_{C,t}$.
\end{enumerate}
\end{assumption}

\begin{assumption}[\textbf{Cross row (column) correlation of noise $\bE_t$}] \label{assume:cross_row_col_corr}
There exists some positive constant $C < \infty$ such that,
\begin{enumerate}[label=(\alph*)]

\item Let  $\bU_E = \E{ \bE_t \bE_t^T/q}$ and $\bV_E = \E{\bE_t^T \bE_t/p}$, we assume $\left\Vert{ \bU_E }\right\Vert_1 \le C$ and $\left\Vert{ \bV_E }\right\Vert_1 \le C$.

\item For all $i \in [p]$ and $j \in [q]$ and $t \in [T]$, we assume $
\sum_{\substack{l\in [p] \\ l\ne i}} \sum_{\substack{h\in [q] \\ h\ne j}} \abs{\E{e_{t,ij} e_{t,lh}}} \le C$.

\item For any row $i, l \in[p]$, any time $t, s \in [T]$, and any column $j\in [q]$,
\begin{equation*}
\sum_{m\in[p]}
\sum_{\substack{h\in [q] \\ h\ne j}}
\lvert \Cov{\left(e_{t, ij} e_{t, lj}, \; e_{s, ih} e_{s, mh} \right)} \rvert  \le C
\end{equation*}
Similar, for any column $j, h\in[q]$, any time $t \in [T]$, and any row $i \in [p]$,
\begin{equation*}
\sum_{m \in [q]}
\sum_{s \in [T]}
\sum_{\substack{l \in [p] \\ l \ne i}}
\abs{\Cov{\left( e_{t, ij}e_{t, ih}, \; e_{s, lj} e_{s, lm} \right)} } \le C.
\end{equation*}

\end{enumerate}
\end{assumption}

\begin{assumption}  \label{assume:CLT}
$\bR_{i \cdot}(t/T)$ and $\bC_{j \cdot}(t/T)$ have continuous derivatives up to the second order. Moreover, there exists $m > 2$, $1 < a, b < \infty$, $1/a + 1/b = 1$, $c = 0,1,2$ such that, for some positive $C < \infty$, and any $t\in [T]$,
\begin{enumerate}[label=(\alph*)]
\item For any $l\in [k]$, and $i\in[p]$, $\E{ \abs{\frac{1}{\sqrt{q}} \sum_{j=1}^q e_{t, i j}}^{mb} } = \bigO{1}$, $\E{ \norm{\frac{1}{\sqrt{q}} \sum_{j=1}^q \bC_{t,j \cdot}^{\left(c\right)} e_{t, i j}}^{mb} } = \bigO{1}$, and $\max_t\E{\norm{\bff_{t, l \cdot}}^{ma}} \le C$.
\item For any $h \in [r]$, and $j \in[q]$,
$\E{ \abs{\frac{1}{\sqrt{p}} \sum_{i=1}^p e_{t, i j}}^{mb} } = \bigO{1}$,
$\E{ \norm{\frac{1}{\sqrt{p}} \sum_{i=1}^p \bR_{t,i \cdot}^{\left(c\right)}  e_{t, i j}}^{mb} } = \bigO{1}$,
and $\max_t\E{\norm{\bff_{t, \cdot h}}^{ma}} \le C$.
\item $\E{\abs{\frac{1}{\sqrt{pq}} \sum_{i=1}^p \sum_{j=1}^q e_{t, i j}}^{mb} } = \bigO{1}$ and $\E{\norm{\frac{1}{\sqrt{pq}} \sum_{i=1}^p \sum_{j=1}^q \bR_{t,i \cdot}^{\left(c\right)}\bC_{t,j \cdot}^{\left(c\right)\top} e_{t, i j}}^{mb} } = \bigO{1}$.
\end{enumerate}
\end{assumption}

\begin{assumption}  \label{assume:kernel}
$k: [-1,1] \rightarrow \mathbb{R}%
^+$ is a symmetric and Lipschitz
continuous probability density function such that  $\int_{-1}^1 k(u) du = 1$, $\int_{-1}^1 uk(u) du = 0$, and $\int_{-1}^1 u^2k(u) du < \infty$.
\end{assumption}

The $\alpha$-mixing condition in Assumption \ref{assume:alphamixing} allows weak temporal correlations for both the factors and noises.  Assumption \ref{assume:factor-noise-matrix} mainly imposes moment conditions on the errors, factors, and their interactions. Assumption \ref{assume:loading_matrix} is an extension of the pervasive assumption \citep{stock2002forecasting} to the matrix variate data.
It ensures that each row and column of the factor matrix $\bF_t$ has a nontrivial contribution to the variance of rows and columns of $\bY_t$. Assumptions \ref{assume:alphamixing}-\ref{assume:loading_matrix} are standard in the literature on factor models and similar assumptions have been imposed in \cite{chen2021statistical} and \cite{yu2021projection}. 

Assumption \ref{assume:alphamixing} addresses temporal dependence, while the dimensions $p$ and $q$ influence our estimator's convergence rates through cross-row and cross-column dependencies. To leverage information across rows and columns, we introduce Assumptions \ref{assume:cross_row_col_corr} and \ref{assume:CLT}, which are met if errors $\bE_t$ are i.i.d. across rows and columns at any time $t$, independent from factor $\bF_t$, and meet specific moment conditions. These assumptions accommodate mild cross-row, cross-column, and temporal dependencies. Assumption \ref{assume:kernel} is standard for kernel regression and is fulfilled by typical second-order kernels like Epanechnikov, uniform, and quartic kernels.

Let $\hat{\bV}_{R,t} \in \mathbb{R}^{k\times k}$ and $\hat{\bV}_{C,t} \in \mathbb{R}^{r\times r}$ be the time-varying diagonal matrices consisting of the first $k$ and $r$ largest eigenvalues of $\hat{\bM}_{R,t}$ ($\hat{\bM}_{C,t}$) in a descending order respectively. By the definition of our estimators $\hat{\bR_t}$ and $\hat{\bC_t}$, we have 
\[
\hat{\bR}_t = \hat{\bM}_{R,t}\hat{\bR}_t\hat{\bV}_{R,t}^{-1} \mbox{\ \ and \ \ }
\hat{\bC}_t  = \hat{\bM}_{C,t}\hat{\bC}_t\hat{\bV}_{C,t}^{-1}. 
\]
Similar to traditional factor models, the latent factor matrix $\bF_t$ and loading matrices $\bR_t$ and $\bC_t$ cannot be identified separately but can be estimated through an invertible matrix transformation. There are invertible matrices $\bH_{R,t}$ and $\bH_{C,t}$ making $\hat{\bR_t}$ and $\hat{\bC_t}$ consistent estimators of $\bR_t\bH_{R,t}$ and $\bC_t\bH_{C,t}$, respectively.
\begin{equation}\label{eqn:H_C_t}
\bH_{R,t} = \frac{1}{Tpq} \sum_{s=1}^{T} K_{h,ts}\bF_s\bC_t^\top\bC_t\bF_s^\top\bR_t^\top\hat{\bR}_t\hat{\bV}_{R,t}^{-1}, \quad\text{and}\quad
\bH_{C,t} = \frac{1}{Tpq} \sum_{s=1}^{T} K_{h,ts}\bF_s\bR_t^\top\bR_t\bF_s^\top\bC_t^\top\hat{\bC}_t\hat{\bV}_{C,t}^{-1} .
\end{equation}

\begin{theorem}
\label{consistency} Suppose Assumptions \ref{assume:alphamixing}-\ref{assume:kernel} are satisfied. As $k,r$ fixed, $h\rightarrow
0 $, $Th^3\rightarrow
\infty $,  and  $T,p,q\rightarrow
\infty $, we have 
\begin{equation}\nonumber
\frac{1}{p}\norm{\hat{\bR_t}-\bR_t\bH_{R,t}}_F^2  =  \Op{\frac{1}{p^2}+\frac{1}{Tqh}+h^4}, \quad\text{and}\quad
\frac{1}{q}\norm{\hat{\bC_t}-\bC_t\bH_{C,t}}_F^2  =  \Op{\frac{1}{q^2}+\frac{1}{Tph}+h^4}.
\end{equation}
Consequently, 
\begin{equation}\nonumber
\frac{1}{p}\norm{\hat{\bR_t}-\bR\bH_{R,t}}^2  =  \Op{\frac{1}{p^2}+\frac{1}{Tqh}+h^4}, \quad\text{and}\quad
\frac{1}{q}\norm{\hat{\bC_t}-\bC\bH_{C,t}}^2  =  \Op{\frac{1}{q^2}+\frac{1}{Tph}+h^4}.
\end{equation}
\end{theorem}
Theorem \ref{consistency} details the convergence rates of our nonparametric estimators, $\hat{\bR_t}$ and $\hat{\bC_t}$. By leveraging the local vector factor model approach to vectorized data (\cite{su2017time}), we explore the dynamics of matrix-valued data with:
\begin{equation} \label{eqn:mfm-vec}
\vect{\bY_t} = \bXi_t\vect{\bF_t} + \vect{\bE_t},
\end{equation}
where $\vect{\bY_t} \in \mathbb{R}^{pq}$ and $\vect{\bF_t} \in \mathbb{R}^{kr}$, respectively.
This method, while bypassing the tensor structure in the loading matrix, leads to convergence rates of $\frac{1}{pq} \norm{\hat\bXi_t-\bXi_t\bH_t}^2= \Op{\frac{1}{p^2q^2}+\frac{1}{Th}+h^4}$, with $\bH_t$ being an orthonormal matrix\footnote{The bias terms are missing in \cite{su2017time} but have been corrected in \cite{su2020correction}.}. However, modeling $\vect{\bY_t}$ as $( \bC _t\otimes \bR_t ) \cdot \vect{\bF_t} + \vect{\bE_t}$ within our specific framework offers improved convergence rates due to its unique structure. Estimation of $\bR_t$ and $\bC_t$ from $\bXi_t$ using Kronecker product decomposition introduces potential inaccuracies \citep{cai2019kopa}. 

To analyze the limiting distribution of the nonparametric estimators $\hat{\bR_t}$ and $\hat{\bC_t}$, we introduce
\begin{equation}\label{eqn:sigma_FC}
\bSigma_{FR,t}^{\left(c_1,c_2\right)} = \bE \left( \bF_t^\top \frac{{\bR_t^{\left(c_1\right)\intercal}}\bR_t^{\left(c_2\right)}}{p}\bF_t\right), 
\mbox{ \ \ and \ \ } \bSigma_{FC,t}^{\left(c_1,c_2\right)} = \bE \left( \bF_t \frac{{\bC_t^{\left(c_1\right)\intercal}}\bC_t^{\left(c_2\right)}}{q}\bF_t^\top\right),
\end{equation}
and 
\begin{equation*}
\bSigma_{R,t}^{\left(c\right)} = \frac{\bR_t^\top\bR_t^{\left(c\right)}}{p}, \mbox{\ \ and \ \ } \bSigma_{C,t}^{\left(c\right)} = \frac{\bC_t^\top\bC_t^{\left(c\right)}}{q},
\end{equation*}
where $c_1$, $c_2$ and $c = 0,1,2$, $\bSigma_{FR,t}^{\left(0,0\right)} =  \bSigma_{FR,t}$ and $\bSigma_{FC,t}^{\left(0,0\right)} =  \bSigma_{FC,t}$. 

\begin{assumption} \label{assume:eigen-val}
The matrices $\bOmega_{R,t}\bSigma_{FC,t}$ and $\bOmega_{C,t}\bSigma_{FR,t}$ are positive definite with distinct eigenvalues for each $t$. 
\end{assumption}

\begin{theorem}
\label{asymptotic normality} Under Assumptions \ref{assume:alphamixing}-\ref{assume:eigen-val}, as $k,r$ fixed, $h\rightarrow 0$, $Th^3\rightarrow\infty$,  and  $T,p,q\rightarrow \infty$, 
\begin{itemize}
\item [(i)] For row loading $\bR_t$, if $\sqrt{qTh}/p \rightarrow 0$, for each $i\in[p]$ and $t\in[T]$, we have 
\begin{equation}\nonumber
\sqrt{qTh} \left(  \hat{\bR}_{t,i.} - \bH_{R,t}^\top \bR_{t,i.} - \bB_{R_{t,i}}\right) \rightarrow^{d} N(0,\nu_{0}\bSigma_{R_{t,i}}),
\end{equation}
where
\begin{equation}\nonumber
\bB_{R_{t,i}} = \frac{1}{2}\bV_{R,t}^{-1}h^{2}\mu_{2}\sum_{c_1+c_2+c_3+c_4=2}^{}
\bH_{R,t}^\top \bSigma_{R,t}^{\left(c_1\right)}\bSigma_{FC,t}^{\left(c_2,c_3\right)}
\bR_{t,i.}^{\left(c_4\right)}+o_{P}\left(  h^{2}\right),
\end{equation}
\begin{equation}\nonumber
\bSigma_{R_{t,i}} =\bV_{R,t}^{-1}\bQ_{R,t}\bPhi_{R,t,i}\bQ_{R,t}^\top\bV_{R,t}^{-1},
\end{equation}
\begin{equation} \label{eqn:phi}
\begin{aligned}
	\bPhi_{R, t,i} & = \underset{q, T \rightarrow \infty}{{\rm p lim}} \frac{1}{qT}  \sum_{r=1}^{T} \sum_{s=1}^{T} \E{\bF_r \bC_t^\top \be_{r, i \cdot} \be_{s, i \cdot}^\top \bC_t \bF_s^\top}, \\
\end{aligned}
\end{equation}
$\nu _{0}=\int_{-1}^{1}k^{2}(u)du$, $\mu _{2}=\int_{-1}^{1} u^{2}k(u)du$, $\bQ_{R,t} = \bV_{R,t}^{1/2}\bPsi_{R,t}^\top\bSigma_{FC,t}^{- 1/2}$, $\bV_{R,t}$ is a diagonal matrix whose entries are the eigenvalues of $\bSigma_{FC,t}$ in decreasing order,  $\Psi_{R,t}$ is the corresponding eigenvector matrix such that $\Psi_{R,t}^\top \Psi_{R,t} = \bI$, and $\bSigma_{FC,t}$ is defined in \eqref{eqn:sigma_FC}.
\end{itemize}

\begin{itemize}
\item [(ii)] For column row loading $\bC_t$, if $\sqrt{pTh}/q \rightarrow 0$, for each $j\in[q]$ and $t\in[T]$, we have
\begin{equation}\nonumber
\sqrt{pTh} \left(  \hat{\bC}_{t,j.} - \bH_{C,t}^\top \bC_{t,j.} - \bB_{C_{t,j}}\right) \rightarrow^{d} N(0,\nu_0\bSigma_{C_{t,j}}),
\end{equation}
where
\begin{equation}\nonumber
\bB_{C_{t,j}} = \frac{1}{2}\bV_{C,t}^{-1}h^{2}\mu_{2}\sum_{c_1+c_2+c_3+c_4=2}^{}
\bH_{C,t}^\top \bSigma_{C,t}^{\left(c_1\right)}\bSigma_{FR,t}^{\left(c_2,c_3\right)}
\bC_{t,j.}^{\left(c_4\right)}+o_{P}\left(  h^{2}\right),
\end{equation}
and
\begin{equation}\label{eqn:phic}
\bSigma_{C_{t,j}} =\bV_{C,t}^{-1}\bQ_{C,t}\bPhi_{C,t,j}\bQ_{C,t}^\top\bV_{C,t}^{-1},
\end{equation}
$\bPhi_{C,t,j}$ is defined similarly to $\bPhi_{R,t,i},$ $\bQ_{C,t} = \bV_{C,t}^{1/2}\Psi_{C,t}^\top\bSigma_{FR,t}^{- 1/2}$, $\bV_{C,t}$ is a diagonal matrix whose entries are the eigenvalues of $\bSigma_{FR,t}$ in decreasing order,  $\Psi_{C,t}$ is the corresponding eigenvector matrix such that $\Psi_{C,t}^\top \Psi_{C,t} = \bI$, and $\bSigma_{FR,t}$ is defined in \eqref{eqn:sigma_FC}.
\end{itemize}
\end{theorem}

Theorem \ref{asymptotic normality} establishes the asymptotic normality of $\hat{\bR_t}$ and $\hat{\bC_t}$ estimators. Both in the interior and boundary regions, the asymptotic bias $\bB_{R_{t,i}}$ and $\bB_{C_{t,j}}$ is linked to $h^2$, attributed to employing a boundary-adjusted kernel function as specified in \eqref{eqn:boundary_kernel}. The pointwise asymptotic mean squared error (AMSE) of $\hat{\bR}_{t,i.}$ can be easily determined by
\begin{align}
AMSE_R& =\frac{h^4\mu_2^2}{4} \norm{\sum_{c_1+c_2+c_3+c_4=2}^{}
\bV_{R,t}^{-1}\bH_{R,t}^\top \bSigma_{R,t}^{\left(c_1\right)}\bSigma_{FC,t}^{\left(c_2,c_3\right)}
\bR_{t,i.}^{\left(c_4\right)}}^2+\frac{\Tr{\nu_0\bSigma_{R_{t,i}}}}{qTh} ,
\end{align}%
and hence the optimal bandwidth for $\hat{\bR}_{t,i.}$ is
\begin{equation}
h_R^{opt}=(qT)^{-\frac{1}{5}}\bigg(\frac{\Tr{\nu_0\bSigma_{R_{t,i}}}}{\mu _{2}^{2}\norm{\sum_{c_1+c_2+c_3+c_4=2}^{}
\bV_{R,t}^{-1}\bH_{R,t}^\top \bSigma_{R,t}^{\left(c_1\right)}\bSigma_{FC,t}^{\left(c_2,c_3\right)}
\bR_{t,i.}^{\left(c_4\right)}}^2}\bigg)^{\frac{1}{5}}. 
\end{equation}
The optimal bandwidth for $\hat{\bC}_{t,i.}$ can be derived similarly. 
\begin{theorem}
\label{consistency_signal} Suppose Assumptions \ref{assume:alphamixing}-\ref{assume:eigen-val} are satisfied. As $k,r$ fixed, $h\rightarrow
0 $, $Th^3\rightarrow
\infty $,  and  $T,p,q\rightarrow
\infty $, we have 
\begin{equation}\nonumber
\begin{aligned}
& \hat{\bF}_t - \bH_{R,t}^{-1}\bF_t\bH_{C,t}^{-1} = \Op{\frac{1}{\min \left( p,q \right)}+h^2},\\
& \hat{\bS}_{t,ij} - \bS_{t,ij} = \Op{\frac{1}{\min \left( p,q,\sqrt{pTh},\sqrt{qTh} \right)}+h^2},\\
\end{aligned}
\end{equation}
for any $1 \le i \le p$ and $1 \le j \le q$.
\end{theorem}
Theorem \ref{consistency_signal} derives the convergence rate of the estimated latent factor $\hat{\bF_t}$ and signal $\hat{\bS}_{t}$. 
In order to achieve consistency, we require dimensions $p$ and $q$ approach infinity. That is because we need sufficient information accumulation to distinguish the signal $\bS_{t}$ from the noise $\bE_t$ at each time point $t$. Theorems \ref{asymptotic normality} and \ref{consistency_signal} present the asymptotic properties when the dimension of the latent matrix factor $k \times r$ is assumed to be known. In practice we can estimate $k$ and $r$ via the generalized eigenvalue ratio-based estimators defined in \eqref{eqn:eigen_ratio}. The asymptotic validity of $\hat{k}$ and $\hat{r}$ is justified in Theorem \ref{consistency_ratio} below. 
\begin{theorem}
\label{consistency_ratio} Suppose Assumptions \ref{assume:alphamixing}-\ref{assume:eigen-val} are satisfied and $k_{max}$ is a predetermined constant no smaller than $k$ or $r$. Then 
\begin{equation*} 
P(\hat{k}=k) \rightarrow 1
\mbox{\ \ and \ \ } 
P(\hat{r}=r) \rightarrow 1
\end{equation*}
as $h\rightarrow
0 $, $Th^3\rightarrow
\infty $, $T,p,q\rightarrow \infty $, where $\hat{k}$ and $\hat{r}$ are defined in \eqref{eqn:eigen_ratio}.
\end{theorem}
Theorem \ref{consistency_ratio} derives the consistency of rank estimators $\hat{k}$ and $\hat{r}$ based on the eigenvalue ratios. This can be viewed as a generalization of Theorem 1 of \cite{ahn2013} from constant parameter factor models to time-varying matrix factor models.

\section{Smooth Loading Functions}\label{sec:smooth}

In this section, we present an approach to obtaining a `smooth' estimation of the loading functions $\bR_t = \bR(t/T)$ ($\bC_t = \bC(t/T)$), which is important for interpretation and visualization. 
We present the method for $\bR_t$, and that for $\bC_t$ is similar.

First we note that the estimator
$\hat{\bR}_t$ obtained in Section \ref{sec:estimation} are not necessarily smooth over $t/T$, which is incurred by the fact that $\hat{\bR}_t$ are estimated separately for each time $t$, the inherit rotational ambiguity associated with PCA, and the possible eigenvalue coalescing at time $t-1$ and $t$. 
Specifically, the columns of $\hat{\bR}_t$ are $\sqrt{p}$ multiplied by the top $k$ eigenvectors of $\hat{\bM}_{R,t}$ corresponding to the $k\times 1$ eigenvalue  vector $\hat{\balpha}_t = [\hat{\alpha}_{t,1},\cdots,\hat{\alpha}_{t,r}]^\top$ in descending order.
The matrix $\bM_{R,t}$ and $\hat{\bM}_{R,t}$ are both smooth functions of $t$, under the assumption that $\bR_t$ and $\bC_t$ are smooth.
When $\hat{\bM}_{R,t}$
and $\hat{\bM}_{R,t-1}$ are close, the eigenvectors $\hat{\bR}_t$ and $\hat{\bR}_{t-1}$ and the eigenvalues $\hat{\balpha}_t$ and $\hat{\balpha}_{t-1}$ would be close most of the time, except in two situations: (1) a {\bf sign change} in the eigenvalue/eigenvector when $\hat{\alpha}_{t-1,i}$ and $\hat{\alpha}_{t,i}$ changes signs; and (2) an {\bf order switching} in the eigenvalues when the eigenvector $\hat{\bR}_{\cdot j}((t-1)/T)$ is closer to $\hat{\bR}_{\cdot j+1}(t/T)$ than to $\hat{\bR}_{\cdot j}(t/T)$. This happens when 
the eigenvalues at $t-1$ and $t$ corresponding to the smooth eigenvectors switch 
order from $j$ to $j+1$ and vice versa. We assume the switching only involves two adjacent eigenvalues.

\cite{Motta2023} introduce the $\sqrt{2}$-signed and $\sqrt{2}$-bridged methods for tracking smoothly evolving covariance matrices in independent normal observations. The $\sqrt{2}$-signed technique assigns eigenvector signs at time $t$ based on closeness to those at $t-1$, starting with an order determined by singular values at $t=1$. The $\sqrt{2}$-bridged approach identifies coalescing areas where two eigenvalues nearly interchange in magnitude.

Our setting is different from that considered in 
\cite{Motta2023} as we consider cross-moment
$\bM_{R,t}$ with a potentially underlying 
time varying mean, and there is a strong serial 
dependence induced by the smoothing operation in
the construction of $\hat{\bM}_{R,t}$. Empirical 
study shows that the approach by \cite{Motta2023}
indeed encounters difficulty in our situation. 
We propose an alternative method for identifying coalescing regions, a crucial input for the second stage of the procedure introduced by \cite{Motta2023}. We treat the problem as a {\bf jump point detection} problem under nonparametric smoothing framework. Specifically we use left and right kernels to estimate the underlying smooth functions on the left and right of the potential jump points and use them to construct 
a test statistics that is sensitive to jumps. 

Define the one sided kernel $K^*_{h^*,st}=2h^{*-1} k(\frac{s-t}{Th^*})$,
and correspondingly
\[
\widehat{\bM}_{R,t}^+ =\sum_{s=t}^{t+\lfloor h^*T\rfloor}K^*_{h^*,st}\mathbf{Y}_s\mathbf{Y}_s^\top, \mbox{\ \ and \ \ }
\widehat{\bM}_{R,t}^- =\sum_{s=t-\lfloor h^*T\rfloor}^{s=t-1}K^*_{h^*,st}\mathbf{Y}_s\mathbf{Y}_s^\top.
\]
Let $\hat{\bR}_{i,t}^+/\sqrt{p}$ and $\hat{\bR}_{i,t}^-/\sqrt{p}$ be the $i$-th eigenvectors of $\hat{\bM}_{R,t}^+$
and $\hat{\bM}_{R,t}^-$, respectively. 
Our objective is to detect if there is a switching in the eigenvalues/eigenvectors between $t-1$ and $t$.
Assuming switching only occurs between two adjacent eigenvalues, 
we take a sequential approach and define the 
test statistics, for $i=1,\ldots r-1$: 
\begin{equation} \label{eq:test_i}
T_{i,t}=
\lnorm\widehat{\mathbf{R}}_{i,t}^-\widehat{\mathbf{R}}_{i,t}^{-\top}-\widehat{\mathbf{R}}_{i,t}^+
\widehat{\mathbf{R}}_{i,t}^{+\top}\rnorm^2
+
\lnorm\widehat{\mathbf{R}}_{i+1,t}^-\widehat{\mathbf{R}}_{i+1,t}^{-\top}-\widehat{\mathbf{R}}_{i+1,t}^+
\widehat{\mathbf{R}}_{i+1,t}^{+\top}\rnorm^2 -
\lnorm\widehat{\mathbf{R}}_{i:i+1,t}^-\widehat{\mathbf{R}}_{i:i+1,t}^{-\top}-\widehat{\mathbf{R}}_{i:i+1,t}^+\widehat{\mathbf{R}}_{i:i+1,t}^{+\top}\rnorm^2, 
\end{equation}
where $\widehat{\mathbf{R}}_{j:j+1,t}\in \mathrm{R}^{p\times2}$ is a submatrix of $\widehat{\mathbf{R}}_{t}$ composed of its $i$-th and $i+1$-th columns only.
The use of {$\mathbf{R}_{j,t}^\pm{\mathbf{R}_{j,t}^\pm}^\top$} instead of $\mathbf{R}_{j,t}^\pm$ serves the purpose of circumventing the sign switching and also enhancing the differences.
The intuition of the last term in $T_{j,t}$ is 
\vspace{-2ex}
\begin{equation*}
\lnorm\mathbf{R}_{t}^-{\mathbf{R}_{t}^-}^\top-\mathbf{R}_{t}^+{\mathbf{R}_{t}^+}^\top\rnorm^2=\lnorm\sum_{j=1}^r\left(\mathbf{R}_{j,t}^-{\mathbf{R}_{j,t}^-}^\top-\mathbf{R}_{j,t}^+{\mathbf{R}_{j,t}^+}^\top\right)\rnorm^2\leq\sum_{j=1}^r\lnorm\mathbf{R}_{j,t}^-{\mathbf{R}_{j,t}^-}^\top-\mathbf{R}_{j,t}^+{\mathbf{R}_{j,t}^+}^\top\rnorm^2,  
\end{equation*}
with the equality reached when there is no switching.
The last term in \eqref{eq:test_i} actually serves two purposes. First, the last term is not sensitive on whether
there is indeed a switch between $i$-th and $i+1$-th columns, but it tends to capture the time varying differences of the linear space spanned the two columns 
between the left and right side of $t$, hence partially compensating the difference in the first two terms due to the same effect. It makes \eqref{eq:test_i} more sensitive to switching but less sensitive to the time
varying nature of the functions around $t$. Second,
it makes the test statistic less sensitive to the situation when the switching is actually between 
the $i-1$-th and $i$-th eigenvalues/eigenvectors.
In this case, $\lnorm\widehat{\mathbf{R}}_{i,t}^-\widehat{\mathbf{R}}_{i,t}^{-\top}-\widehat{\mathbf{R}}_{i,t}^+
\widehat{\mathbf{R}}_{i,t}^{+\top}\rnorm^2$ will be large, but it may be largely cancelled by the 
last term in \eqref{eq:test_i}.
It is interesting to note that, 
using the orthonormality property of the matrices, 
$T_{j,t}$ can be rewritten as $T_{i,t}=2\left[\left(\widehat{\mathbf{R}}_{i,t}^{-\top}
\widehat{\mathbf{R}}_{i+1,t}^{+}\right)^2
+\left(\widehat{\mathbf{R}}_{i+1,t}^{-\top}
\widehat{\mathbf{R}}_{i,t}^{+}\right)^2\right]$, showing that $T_{i,t}$ actually measures the distance (angle) after swapping the $i$-th and $(i+1)$-th columns.

When there is a switching between $\bR_{i,t}$ 
and $\bR_{i+1,t}$, it has impact on 
their sample versions in a much wider time range, due to the smoothing operation. In addition, when two eigenvalues are the same, their corresponding eigenvectors are not uniquely identified. Hence the estimators $\hat{\bR}_{i,t}$ would be very inaccurate when $t$ is close to the 
actual switching point. Hence corrections and re-estimation are needed in a region around the switching point.
We determine the significant regions (coalescing regions) that are 
impacted by the switching
using the continuous subsets of $T_{i,t}$ that 
exceeds a certain 
thresholds $thr_i$ for $i\in[r-1]$, similar to an outlier detection procedure. We define these coalescing regions as $P_m=(a_m,b_m]\subset[T]$ and its associated switching column $\kappa_m$ such that for all $t\in P_m$, $T_{\kappa_m,t}>thr_{k_m}$ and $T_{\kappa_m,a_m-1}\leq thr_{\kappa_m}$ and $T_{\kappa_m,b_m+1}\leq thr_{\kappa_m}$.
Here, $m=1,2,\cdots,M$ where $M$ is the total number of regions. We use the thresholds $thr_i=\max\{1.5IQR_i, 1.5 IQR\}$, where $IQR_j$ is the interquartile range of $T_{j,t}$ and $IQR$ is the global interquartile range employing the data set $\{T_{j,t}:j\in[r-1];\ t=[T]\}$. It is possible to develop a bootstrap 
type of procedures similar to that of \cite{Motta2023} though it is much more involved due to our complex setting. Empirically we found the {\it ad hoc} procedure using IQR is quite reasonable.

Next, using the coalescing regions $P_m$ and its associated switching order $\kappa_m$, we follow the approach of \cite{Motta2023} to perform a switching and smoothing operation. Specifically, starting from $i=1$, we identify all regions $P_m$'s such that $\kappa_m=i$, perform the switching of $\hat{\bR}_{i,t}$ outside these intervals, and apply smoothing splines to interpolate the functions inside the $P_m$ regions. The interpolated function are further adjusted using Gram-Schmidt orthogonalization to obtain the final estimates \citep{Motta2023}. 

Note that the smoothed loading functions are still subject to {\bf global} rotation ambiguity. To obtained more interpretable representation, we perform a global 
varimax \citep{Kaiser1958} rotation on the $(dT)\times r$ matrix by stacking the smoothed $\tilde{\bR}_{t}$, $t\in[T]$ matrices.

\section{Simulation} \label{sec-simul}
In this section, we study the numerical performance of the proposed time-varying matrix-valued approach. Throughout, the matrix observations $\bY_t$'s are generated according to model~\eqref{eqn:mfm}.
The dimension of the latent factor matrix $\bF_t$ is fixed at $k \times r = 2 \times 2$.
The values of $p$, $q$, and $T$ vary in different settings.
We simulate $\vect{\bF_t}$ from the following Vector Auto-Regressive model of order one (VAR(1) model):
\begin{equation}
\vect{\bF_t} = \bPhi \cdot \vect{\bF_{t-1}} + \bepsilon_t, \label{simu_F}
\end{equation}
where the AR coefficient matrix $\bPhi = 0.1 \cdot \bI_4$ and $\Var{(\bepsilon_t)} = 0.99\cdot\bI_4$. Thus, $\Var{\vect{\bF_t}} = \bI_4$. We simulate noise $\bE_t$ also from VAR(1),
\begin{equation}
\vect{\bE_t} = \bGamma \cdot \vect{\bE_{t-1}} + \bu_t,  \label{simu_E}
\end{equation}
where $\bGamma = \psi \cdot \bI_{pq}$ and $\Var{\bu_t} = (1 - \psi^2)\bI_{pq}$. Thus, $\Var{\vect{\bE_t}} = \bI_{pq}$. We choose $\psi = 0.1$ and then increase to $\psi = 0.5$ to examine how temporal dependence may affect our results. 
The true loading matrices $\bR_t$ and $\bC_t$ are generated as:
\begin{itemize}[before=\itshape,font=\normalfont,leftmargin=*]
\item DGP1: The first column of $\bR_t$: $\bR_{t,.1}=\bR_{.1}$, the second column of $\bR_t$: $\bR_{t,.2}=\bR_{.2}+\bG_t$; the first column of $\bC_t$: $\bC_{t,.1}=\bC_{.1}$, the second column of $\bC_t$: $\bC_{t,.2}=\bC_{.2}+\bH_t$. The time-invariant loading matrices $\bR_{.j}$ and $\bC_{.j}$ are independently sampled from $\calU \left( -1, 1 \right)$ for $j=1,2$, $\bG_{t}=\bG \left(\frac{t}{T} \right)=2 \left(\frac{t}{T}\right)+\exp{\left[-16\left(\frac{t}{T}-0.5\right)^2\right]}-1$ and $\bH_t=0.2\exp{\left[-0.7+3.5(\frac{t}{T})\right]}$. 
\item DGP2: The first column of $\bR_t$: $\bR_{t,.1}=\bR_{.1}+\bG_t$, the second column of $\bR_t$: $R_{t,i2}=\bF\left(10t/T;2;5i/p+2\right)$, $i=1,...,p$; the first column of $\bC_t$: $\bC_{t,.1}=\bC_{.1}+\bH_t$, the second column of $\bC_t$: $C_{t,j2}=\bF\left(10t/T;2;5j/q+2\right)$, $j=1,...,q$. The time-invariant loading matrices $\bR_{.j}$ and $\bC_{.j}$ are independently sampled from $\calN \left( 0, 1 \right)$ for $j=1,2$, $b=2$, and $F\left(\tau;\kappa;\gamma\right)=\{1+\exp [-\kappa(\tau-\gamma)]\}^{-1}$. 
\end{itemize}
We first study the performance of our proposed approach on estimating the time-varying loading matrices $\bR_t$ and $\bC_t$. We consider three pairs of $\left( p,q \right)$ combinations: $\left( 20,20 \right)$, $\left(100,20 \right)$, or $\left( 100, 100 \right)$. The sample size $T$ is selected as $100$, $200$, and $400$, and simulation results are based on $100$ repetitions. For nonparametric estimation, we use the Epanechnikov kernel and Silverman's rule-of-thumb (RoT) bandwidth  $h_R=c(qT)^{-\frac{1}{5}}$ and $h_C=c(pT)^{-\frac{1}{5}}$, where $c=\frac{2.345}{\sqrt{12}}$, for estimating $\bR_t$ and $\bC_t$ respectively. This simple RoT bandwidth attains the optimal rate for local smoothing as shown in Theorem \ref{asymptotic normality}. We also try the RoT bandwidth with different scaling parameters, and the simulation shows that our estimation results are not very sensitive to the bandwidth selection. To evaluate the accuracy of our estimator, we use the average column space distance
\begin{equation} \label{eqn:spdist}
\bar{\calD}_{\left(\hat\bA, \bA\right)} = \frac{1}{T} \sum_{t = 1}^{T}  \norm{\hat\bA_t\left(\hat\bA_t^\top\hat\bA_t\right)^{-1}\hat\bA_t^\top - \bA_t\left(\bA_t^\top\bA_t\right)^{-1}\bA_t^\top}.
\end{equation}
Table \ref{table:estimating_loading} reports the mean and standard deviation of the average column space distance. For both DGPs considered, $\bar{\calD}_{\left(\hat\bR, \bR\right)}$ and $\bar{\calD}_{\left(\hat\bC, \bC\right)}$ decrease with the increase
of $p,q$ and $T$. The estimation results are robust to the the temporal dependence of noise $\bE_t$.  

\begin{table}[ht!]
\centering
\caption{ Estimation of loading matrices }
\resizebox{0.8\linewidth}{!}{
\begin{tabular}{c| c c c c c c c } \hline
&  &  \multicolumn{2}{c}{$T = 100$} & \multicolumn{2}{c}{$T = 200$} & \multicolumn{2}{c}{$T = 400$}  \\ \hline
$DGP$ & $(p,q)$  & $\bar{\calD}_{\left(\hat\bR, \bR\right)}$ &  $\bar{\calD}_{\left(\hat\bC, \bC\right)}$ & $\bar{\calD}_{\left(\hat\bR, \bR\right)}$ &  $\bar{\calD}_{\left(\hat\bC, \bC\right)}$ & $\bar{\calD}_{\left(\hat\bR, \bR\right)}$ &  $\bar{\calD}_{\left(\hat\bC, \bC\right)}$ \\ \hline
\multirow{8}{*}{$1$}  &  \multicolumn{7}{c}{$\psi=0.1$}\\ 
& ($20,20$) & 1.09(0.21) & 0.98(0.15)  & 0.98(0.12) & 0.55(0.07) & 0.58(0.08) & 0.50(0.04) \\
& ($100,20$) & 0.98(0.12) & 0.55(0.07)  & 0.73(0.08) & 0.40(0.05) & 0.41(0.04) & 0.35(0.02) \\
& ($100,100$) & 0.58(0.08) & 0.50(0.04)  & 0.52(0.04) & 0.29(0.04) & 0.29(0.02) & 0.26(0.02) \\
& \multicolumn{7}{c}{$\psi=0.5$}\\ 
& ($20,20$) & 1.14(0.22) & 1.04(0.17)  & 1.01(0.12) & 0.58(0.08) & 0.59(0.08) & 0.52(0.04) \\
& ($100,20$) & 1.01(0.12) & 0.58(0.08)  & 0.75(0.08) & 0.42(0.06) & 0.42(0.04) & 0.37(0.02) \\
& ($100,100$) & 0.59(0.08) & 0.52(0.04)  & 0.54(0.05) & 0.30(0.04) & 0.30(0.03) & 0.27(0.02) \\\hline 
\multirow{8}{*}{$2$}  &  \multicolumn{7}{c}{$\psi=0.1$}\\ 
& ($20,20$) & 2.46(0.20) & 2.15(0.20)  & 2.27(0.18) & 1.94(0.16) & 2.11(0.14) & 1.79(0.16) \\
& ($100,20$) & 2.35(0.20) & 1.72(0.17)  & 2.14(0.14) & 1.57(0.13) & 1.95(0.11) & 1.42(0.10) \\
& ($100,100$) & 1.91(0.20) & 1.47(0.13)  & 1.73(0.13) & 1.29(0.08) & 1.61(0.10) & 1.15(0.09) \\
& \multicolumn{7}{c}{$\psi=0.5$}\\ 
& ($20,20$) & 2.48(0.20) & 2.24(0.20)  & 2.30(0.18) & 2.05(0.15) & 2.13(0.14) & 1.88(0.16) \\
& ($100,20$) & 2.38(0.19) & 1.80(0.17)  & 2.17(0.14) & 1.64(0.13) & 1.97(0.11) & 1.50(0.11) \\
& (100,100) & 1.93(0.19) & 1.55(0.13)  & 1.75(0.13) & 1.37(0.08) & 1.62(0.10) & 1.23(0.09) \\\hline 
\end{tabular}}
\begin{tablenotes}
\small
\item Note: Mean and standard deviation in parentheses of $\bar{\calD}_{\left(\hat\bR, \bR\right)}$ and $\bar{\calD}_{\left(\hat\bC, \bC\right)}$ from 100 iterations. 
\end{tablenotes}
\label{table:estimating_loading}
\end{table}

With the same simulated data, we compare the proposed time-varying matrix-valued approach and the time-varying vector-valued approach in \cite{su2017time} through the estimation accuracy of the total loading matrix $\bXi_t=\bC _t\otimes \bR_t$. In what follows, the subscripts
mat and vec denote our approach and  \cite{su2017time}’s method, respectively. The loading space $\hat{\bXi}_{mat,t}$ is computed as
$\hat{\bXi}_{mat,t}=\hat{\bC _t}\otimes \hat{\bR_t}$, where $\hat{\bC _t}$ and $\hat{\bR _t}$ are our nonparametric estimators. For the vector-valued approach, we apply \cite{su2017time}’s method to the vectorized observations ${\vect{\bY_t}, t = 1, 2,\cdots, T}$ as in \eqref{eqn:mfm-vec} to obtain $\hat{\bXi}_{vec,t}$. The results
for the estimation accuracy of $\bXi_t$ measured by $\bar{\calD}_{mat}$ and $\bar{\calD}_{vec}$ are reported in Table \ref{table:estimating_comparison}, which shows that the matrix approach efficiently improves the estimation accuracy over the vector-valued approach. We only consider three pairs of small $\left( p,q \right)$ combinations: $\left( 10,10 \right)$, $\left( 20,10 \right)$, or $\left( 20, 20 \right)$ because the estimation of time-varying vectorized factor models is extremely time-consuming. Nevertheless we expect the same pattern of estimation comparison would continue to hold for large $p$ and $q$.   

\begin{table}[ht!]
\centering
\caption{ Comparison of estimation accuracy }
\resizebox{0.6\linewidth}{!}{
\begin{tabular}{c| c c c c c } \hline
&  &   \multicolumn{2}{c}{$T = 100$} & \multicolumn{2}{c}{$T = 200$}  \\ \hline
$DGP$ & $(p,q)$  & $\bar{\calD}_{max}$ &  $\bar{\calD}_{vec}$ & $\bar{\calD}_{max}$ &  $\bar{\calD}_{vec}$ \\ \hline
\multirow{8}{*}{$1$}  &  \multicolumn{5}{c}{$\psi=0.1$}\\ 
& ($10,10$) & 0.23(0.06) & 0.78(0.10)  & 0.13(0.02) & 0.62(0.10)  \\
& ($20,10$) & 0.17(0.04) & 0.65(0.11)  & 0.15(0.02) & 0.75(0.07) \\
& ($20,20$) & 0.19(0.03) & 0.76(0.09)  & 0.11(0.02) & 0.60(0.08) \\
& \multicolumn{5}{c}{$\psi=0.5$}\\ 
& ($10,10$) & 0.25(0.07) & 0.90(0.09)  & 0.14(0.03) & 0.79(0.12)  \\
& ($20,10$) & 0.18(0.05) & 0.79(0.13)  & 0.16(0.03) & 0.91(0.06) \\
& ($20,20$) & 0.20(0.04) & 0.91(0.06)  & 0.12(0.02) & 0.78(0.12) \\ \hline 
\multirow{8}{*}{$2$}  &  \multicolumn{5}{c}{$\psi=0.1$}\\ 
& ($10,10$)  & 0.34(0.03) & 0.66(0.11)  & 0.31(0.02) & 0.58(0.12)  \\
& ($20,10$)  & 0.32(0.02) & 0.64(0.10)  & 0.29(0.01) & 0.54(0.08) \\
& ($20,20$) & 0.29(0.02) & 0.61(0.09)  & 0.27(0.02) & 0.51(0.07) \\
& \multicolumn{5}{c}{$\psi=0.5$}\\ 
& ($10,10$) & 0.35(0.03) & 0.70(0.12)  & 0.32(0.03) & 0.61(0.13)  \\
& ($20,10$) & 0.33(0.02) & 0.69(0.11)  & 0.29(0.02) & 0.58(0.09) \\
& ($20,20$) & 0.30(0.02) & 0.65(0.10)  & 0.27(0.02) & 0.55(0.09) \\
\hline 
\end{tabular}%
}
\begin{tablenotes}
\small
\item Note: Mean and standard deviation in parentheses of $\bar{\calD}_{max}$ and $\bar{\calD}_{vec}$ from 100 iterations. 
\end{tablenotes}
\label{table:estimating_comparison}
\end{table}

We next evaluate the performance of the matrix-valued approach on estimating the number of factors $k$ and $r$. Table \ref{table:estimating_k_r} shows the relative frequencies of estimated rank pairs over 100 iterations. The three pairs $(2, 1), (1, 2)$ and $(2,2)$ have high appearances in all of the combinations of $(p,q,T)$. The row for the true rank pair $(2, 2)$ is highlighted. It shows that the relative frequency of correctly estimating the true rank pair increases with the increase of $(p,q,T)$ and approaches unity. 
\begin{table}[htpb!]
\centering
\caption{ Estimation of latent dimensions } \label{table:estimating_k_r}
\resizebox{0.88\linewidth}{!}{
\begin{tabular}{c|c|c c c c c c c c c} 
\hline
&  & \multicolumn{3}{c}{$p,q = 20,20$} & \multicolumn{3}{c}{$p,q = 100,20$} & \multicolumn{3}{c}{$p,q = 100,100$}  \\ \hline
$DGP$ & $(\hat{k},\hat{r})$  & $T=100$ &  $T=200$ & $T=400$ &  $T=100$ &  $T=200$ & $T=400$ &
$T=100$ &  $T=200$ & $T=400$   \\ \hline
\multirow{10}{*}{$1$}  &  &\multicolumn{9}{c}{$\psi=0.1$} \\ 
& ($2,1$) & 0.13 & 0.06  & 0.04 & 0.02 & 0.03 & 0.01 & 0.00 & 0.00 & 0.00\\
& ($1,2$) & 0.05 & 0.02  & 0.03 & 0.00 & 0.00 & 0.00 & 0.00 & 0.00 & 0.00 \\
\rowcolor[HTML]{EFEFEF} 
& ($2,2$) & 0.80 & 0.90  & 0.93 & 0.98 & 0.97 & 0.99 & 1.00 & 1.00 & 1.00\\
& other & 0.02 & 0.01  & 0.00 & 0.00 & 0.00 & 0.00 & 0.00 & 0.00 & 0.00\\
& & \multicolumn{9}{c}{$\psi=0.5$}\\ 
& ($2,1$) & 0.12 & 0.08  & 0.03 & 0.02 & 0.02 & 0.00 & 0.00 & 0.00 & 0.00\\
& ($1,2$) & 0.09 & 0.04  & 0.03 & 0.00 & 0.00 & 0.00 & 0.00 & 0.00 & 0.00 \\
\rowcolor[HTML]{EFEFEF}
& ($2,2$) & 0.77 & 0.87  & 0.94 & 0.98 & 0.98 & 1.00 & 1.00 & 1.00 & 1.00\\
& other & 0.02 & 0.01  & 0.00 & 0.00 & 0.00 & 0.00 & 0.00 & 0.00 & 0.00\\ \hline
\multirow{10}{*}{$2$}  & & \multicolumn{9}{c}{$\psi=0.1$}\\ 
& ($2,1$) & 0.21 & 0.16 & 0.14 & 0.21 & 0.11 & 0.06 & 0.00 & 0.00 & 0.00\\
& ($1,2$) & 0.02 & 0.01 & 0.00 & 0.00 & 0.00 & 0.00 & 0.00 & 0.00 & 0.00 \\
\rowcolor[HTML]{EFEFEF} 
& ($2,2$) & 0.74 & 0.81  & 0.86 & 0.79 & 0.89 & 0.94 & 1.00 & 1.00 & 1.00\\
& other & 0.03 & 0.02  & 0.00 & 0.00 & 0.00 & 0.00 & 0.00 & 0.00 & 0.00\\
& & \multicolumn{9}{c}{$\psi=0.5$}\\ 
& ($2,1$) & 0.22 & 0.16  & 0.15 & 0.20 & 0.12 & 0.07 & 0.00 & 0.00 & 0.00\\
& ($1,2$) & 0.02 & 0.02  & 0.00 & 0.00 & 0.00 & 0.00 & 0.00 & 0.00 & 0.00 \\
\rowcolor[HTML]{EFEFEF}
& ($2,2$) & 0.73 & 0.80  & 0.85 & 0.80 & 0.88 & 0.93 & 1.00 & 1.00 & 1.00 \\
& other & 0.03 & 0.02  & 0.00 & 0.00 & 0.00 & 0.00 & 0.00 & 0.00 & 0.00 \\ \hline 
\end{tabular}%
}
\begin{tablenotes}
\small
\item Note: Relative frequency of estimated rank pair $(\hat{k},\hat{r})$ from 100 iterations. The row with the true rank pair $(2,2)$ is highlighted. 
\end{tablenotes}
\end{table}

We evaluated the smoothing operation's empirical performance on the load matrices as described in Section~\ref{sec:smooth}. To closely mirror the switching behavior observed in the real data example in Section~\ref{sec-appl-it}, we generated data for controlled experiments:
\begin{equation} \label{simu}
\mathbf{X}_t = \mathbf{R}_t \mathbf{F}_t \mathbf{C}_t \mbox{ \ and \ \ } 
\mathbf{Y}_t = \alpha \mathbf{X}_t + \mathbf{E}_t,
\end{equation}
where $\bF_t$ and $\bE_t$ follows the simulation setting above in \eqref{simu_F} and \eqref{simu_E}, with a $4\times 4$ $\bF_t$ factor process.
The parameter $\alpha^2$ denotes the signal-to-noise ratio (SNR) (i.e. $(1+\alpha^2)^{-1}$ is the relative size of the noise process with respect to $\mathbf{Y}_t$), after $\bX_t$ is normalized so
$\mathrm{var}(\mathrm{vec}(\mathbf{X}_t)) = \mathrm{var}(\mathrm{vec}(\mathbf{X}_t)) = \mathbf{I}_{pq}$. 

The time-varying matrices $\mathbf{R}_t$ and $\mathbf{C}_t$ are constructed similarly to those in \cite{Motta2023}. Specifically, 
we first generate the time-varying matrix $\mathbf{Q}_t$ whose $j$-th column is 

\begin{equation}
\mathbf{Q}_{\cdot j,t} =
\begin{bmatrix}
\sin\left((j-1)\pi w \frac{t}{T} - \phi\right)\sqrt{2/3}  \\
\cos\left((j-1)\pi w \frac{t}{T} - \phi\right)\sqrt{2/3}\\
\exp\left((j-1)\pi w \frac{t}{T} - \phi\right)\sqrt{2/3}\\
\exp\left(-(j-1)\pi w \frac{t}{T} - \phi\right)\sqrt{2/3}.
\end{bmatrix}
\end{equation} where $w=0.75$ and $\phi=0.2$. 
Let $\overline{\mathbf{R}}_t = \mathbf{U}_{R}\mathbf{Q}_{R,t}\left(\mathbf{Q}_{R,t}^{\top}\mathbf{Q}_{R,t}\right)^{-1/2}$ and $\overline{\mathbf{C}}_t = \mathbf{U}_{C}\mathbf{Q}_{C,t}\left(\mathbf{Q}_{C,t}^{\top}\mathbf{Q}_{R,t}\right)^{-1/2}$, 
where $\mathbf{U}_{R}\in \mathrm{R}^{p\times k}$ and {$\mathbf{U}_{C}\in \mathrm{R}^{q\times r}$} are orthonormal matrices randomly generated.
Then the matrices $\mathbf{R}_t$ and $\mathbf{C}_t$ are generated as $\mathbf{R}_t = \overline{\mathbf{R}}_t\mathbf{\Lambda}^{1/2}_{R,t}$ and $\mathbf{C}_t = \overline{\mathbf{C}}_t\mathbf{\Lambda}^{1/2}_{C,t}$, ensuring that $\overline{\mathbf{R}}_t^\top \overline{\mathbf{R}}_t = \mathbf{I}_k$, $\overline{\mathbf{C}}_t^\top \overline{\mathbf{C}}_t = \mathbf{I}_r$. This normalization is
to ensure that $\alpha^2$ in \eqref{simu} is the true signal-to-noise ratio. 

\begin{figure}[ht]
\centering
\includegraphics[width=0.95\linewidth]{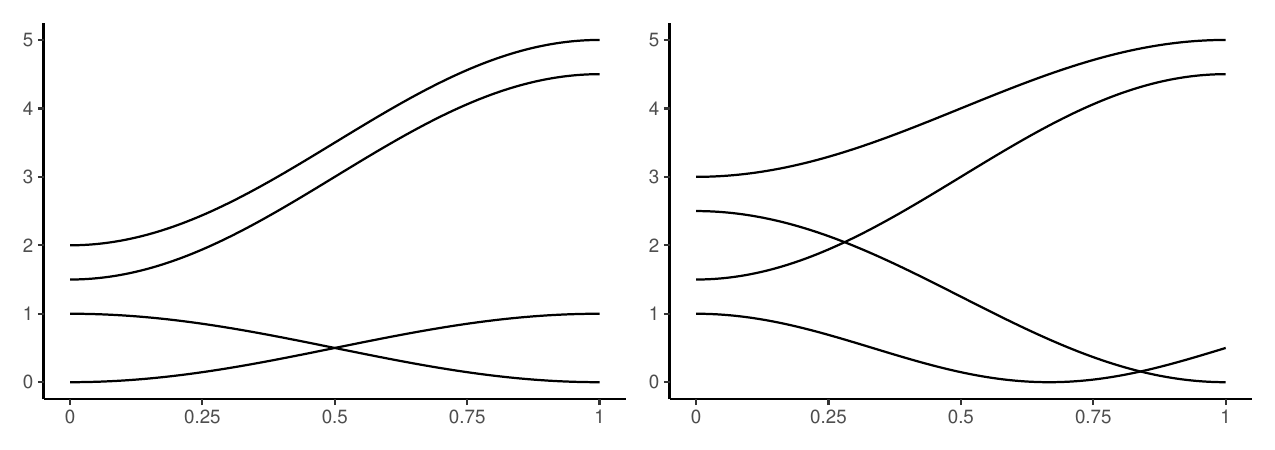}
\caption{Diagonal entries of the matrices $\boldsymbol{\Lambda}_{R,\frac{t}{T}}$ (Left panel) and  $\boldsymbol{\Lambda}_{C,\frac{t}{T}}$ (Right panel) }
\label{fig:Lambda}
\end{figure}

We set the eigenvalues function matrices to explore one and two coalescing points in separate analyzes. Specifically, let 
$\Lambda_{R,t}$ be a $4\times 4$ diagonal matrix, with diagonal elements being 
$3.5-1.5cos(\pi t/T)$,
$3-1.5cos(\pi t/T)$,
$0.5(1+cos(\pi t/T))$, and
$0.5(1-cos(\pi t/T))$. 
Similarly, let $\Lambda_{C,t}$ be a $4\times 4$ diagonal matrix, with diagonal elements being
$4-cos(\pi t/T)$, 
$3-1.5cos(\pi t/T)$,
$1.25(1+cos(\pi t/T))$, and
$0.5+0.5cos(1.5\pi t/T)$.
The functions are shown in Figure \ref{fig:Lambda}.
We show the results of estimating $\bR_t$
and $\bC_t$, using
$\mathbf{X}_t = \mathbf{R}_t \mathbf{F}_t \overline{\mathbf{C}}_t^\top$
and 
$\mathbf{X}_t = \overline{\mathbf{R}}_t \mathbf{F}_t \mathbf{C}_t^\top$, 
to have $E[\mathbf{X}_t^\top\mathbf{X}_t]=k\gamma(0)\mathbf{R}_t\Lambda_{R,t}\mathbf{R}_t^\top$. 
and $E[\mathbf{X}_t\mathbf{X}_t^\top]=r\gamma(0)\mathbf{C}_t\Lambda_{C,t}\mathbf{C}_t^\top$, respectively. 
Here, $\gamma(0)$ is the variance of an AR process with autoregressive parameter $0.1$ and noise variance 1. In this context, we designate these conditions as Scenario 0. 
In Scenario 1 we add a 
deterministic trend to $\bF_t$, using
$\mathbf{F}^{(1)}_t=\mathbf{F}^{(0)}_t+\frac{\sqrt{k}t}{T}\mathbf{M}$ where $\mathbf{M}$ is a randomly generated orthonormal matrix.
In Scenario 2 we simply use 
$\mathbf{F}^{(2)}_t=\frac{t}{T}\mathbf{M}$ as the 
factor process. We are interested in these two 
scenarios because the factor process in our trade volume application is clearly nonstationary, while
the method in \cite{Motta2023} critically rely on 
the stationary processes. In both scenarios the factor processes are scaled to achieve the desired Signal-to-Noise Ratio. 

\begin{figure}[ht]
\centering
\includegraphics[width=\linewidth]{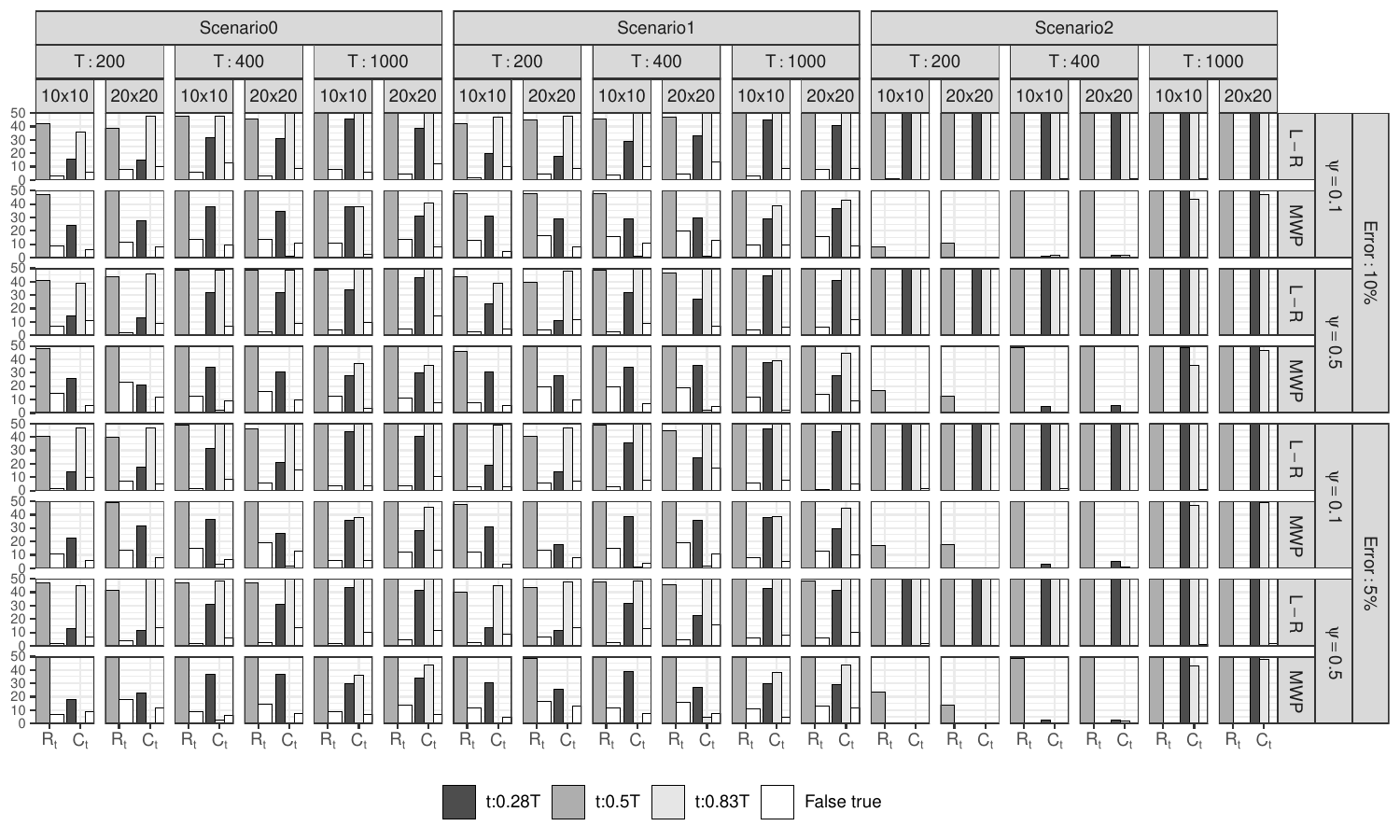}
\caption{Number of times identified coalescing points on each scenario. Varying the dimensions $(p, q) \in {(10, 10), (10, 20), (20, 20)}$, the sample size $T \in {200, 400, 1000}$, the size of the noise $(1+\alpha^2)^{-1} \in \{5\%,10\%\}$, and the AR parameter $\psi \in \{0.1, 0.5\}$. }\label{fig:SimHist}
\end{figure}

We illustrate the procedure by varying the dimensions $(p, q) \in \{(10, 10), (10, 20), (20, 20)\}$, the sample size $T \in \{200, 400, 1000\}$, the error size $(1+\alpha^2)^{-1} \in \{5\%, 10\%\}$, and the AR parameter $\psi \in \{0.1, 0.5\}$.
For each scenario, we obtain 50 replicates and classify the identified regions concerning coalescing points, namely $0.5$ for the eigenvalues of $\bM_{R,t}$ and $\{0.28, 0.83\}$ for that of $\mathbf{M}_{C,t}$. Figure \ref{fig:SimHist} shows the number of times the true coalescing points are in the identified coalescing regions. If an identified coalescing region does not contain any of the true coalescing points, 
it is classified as \textit{"False Positive"}. Within each box, the first two bars are for the 'true positive' and 'false positive' identifications of the single 
coalescing point in $\bM_{R,t}$. The third and fourth bars are for the 'true positive' identifications of the two coalescing point in $\bM_{C,t}$ and the last bar is for the 'false positive' of $\bM_{C,t}$.

In addition, we also obtained the results using 
that in \cite{Motta2023}, where local stationarity is assumed. Here we use the cross-moment $\bM_{R,t}$
and $\bM_{C,t}$, instead of covariance matrix. 
These matrices are then utilized to generate bootstrap samples through the expression $\widehat{\mathbf{R}}t\widehat{\mathbf{\Lambda}}^{1/2}{R,t}\mathbf{Z}_t$, where $\mathbf{Z}_t$ denotes a Gaussian white noise process with zero mean and variance $\mathbf{I}$.  The purpose of developing these bootstrap samples is to delineate the region where eigenvalues coalesce, with 
the 5th percentile as the bounds to identify the coalescing regions. 

Figure \ref{fig:SimHist} shows that both approaches yield an increase in correct identifications and a decrease in false positives as the sample size increases.
Comparing the Left-Right approach (labelled as L-R) and that of \cite{Motta2023} (labelled as MVP)
(e.g. firts row and second row),
it is seen that in relative small sample sizes the left-right 
detection procedure in Section~\ref{sec:smooth} performs better than that of \cite{Motta2023} for $\bC_t$ where there are two switchings. Both procedures are similar when there is only only crossing for $\bR_t$. The procedure of \cite{Motta2023} has difficulty in Scenario 2 where a deterministic trend is present while the left-right detection procedure seems to benefit from the presence of the trend.

\section{Empirical Application} \label{sec-appl-it}

Understanding international trade flow patterns and their evolution is crucial for policy-making, economic forecasting, and firm optimization. International trade data, forming a dynamic sequence of matrix variates, closely mirrors network structures, offering insights into interaction patterns, evolutionary significance, and natural actor groupings within the network. Theoretical and empirical research on international trade networks has advanced significantly \citep{Chaney2016l}, with the Global Vector Autoregressive (GVAR) approach being particularly notable for analyzing trade network interactions \citep{pesaran2016}. For instance, \cite{Bussiere2012} and \cite{Greenwood2012} utilized GVAR models to study demand shocks' effects on global imbalances and forecast trade imbalances across 33 countries, respectively. However, GVAR's reliance on a pre-determined weighting matrix to address dimensionality issues is a limitation. This study introduces a time-varying matrix factor model for international trade data, estimating time-varying loading matrices and latent factors concurrently.

\subsection{Data and sample}

We use monthly multilateral import and export volumes of commodity goods among 24 countries and regions from January 1982 to December 2018. The
data come from the International Monetary Fund Direction of Trade Statistics (IMF-DOTS). The countries and regions included in the study are Australia(AU), Canada(CA), China Mainland(CN), Denmark(DK), Finland(FI), France(FR), Germany(DE), Hong Kong(HK), Indonesia(ID), Ireland(IE), Italy(IT), Japan(JP), Korea(KR), Malaysia(MY), Mexico(MX), Netherlands(NL), New Zealand(NZ), Singapore(SG), Spain(ES), Sweden(SE), Taiwan(TW), Thailand(TH), United Kingdom(GB), and the United States(US). 

In our study, we utilize the time-varying matrix factor model \eqref{eqn:mfm} to examine international trade flow data, where $Y_t$ is depicted as a $24\times24$ matrix, with rows and columns representing exporting and importing countries, respectively. We rely on import data valued in U.S. dollars for their presumed accuracy over export figures \citep{Linnemann1966}. It's noteworthy that Taiwan's trade data, absent in the IMF-DOTS, is inferred from its trading partners' export information. To mitigate the effects of significant one-off transactions and trade anomalies, we calculate trade flows using three-month averages instead of monthly snapshots.

\subsection{Estimation Results}

As described in Section~\ref{sec:model}, the $k\times r$ latent factors in model (\ref{eqn:mfm})  can be interpreted as the transport volumes of a condensed trading network among $k$ export hubs and $r$ import hubs in our application.
We use the generalized ratio-based method in (\ref{eqn:eigen_ratio}) as well as the screen plot to select $\left(4,4\right)$ as the number of dimensions of the latent hubs. For better visualization and interpretation, all estimated $\bR_t$ and $\bC_t$ are normalized so the sum of absolute values of each column is 1.

\begin{figure}[ht!]
\caption{Trading volume example: Initial estimate of $\bR_{t}$.}
\includegraphics[width=1\linewidth,height=1.3in]{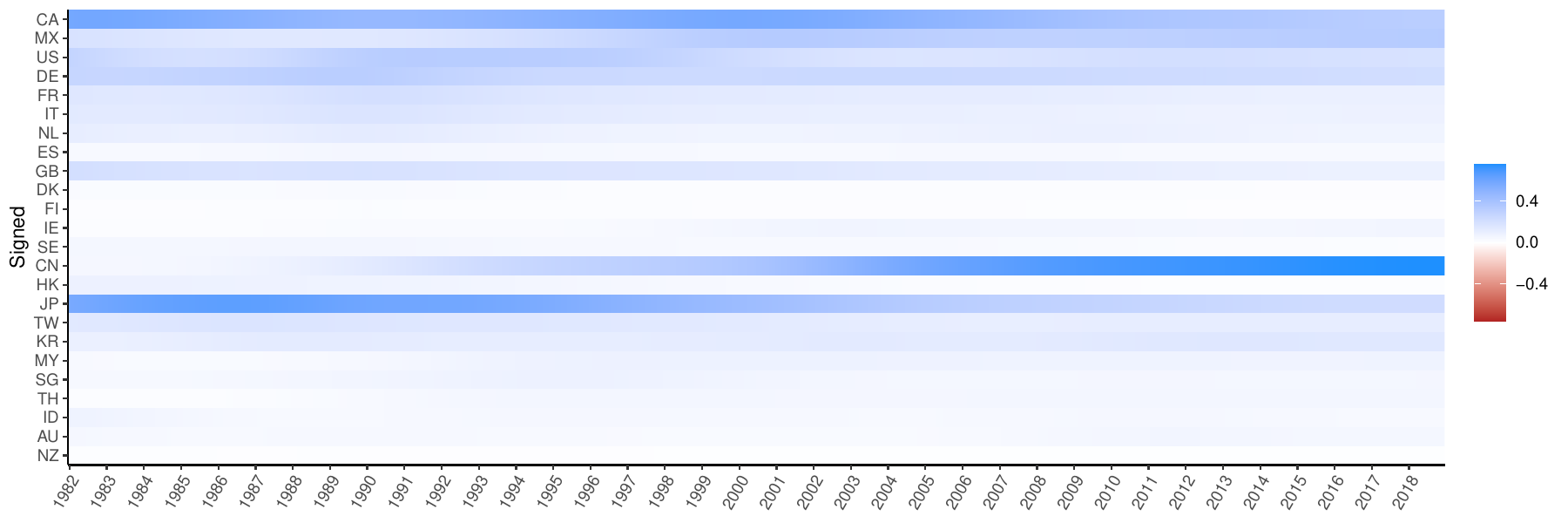}
\includegraphics[width=1\linewidth,height=1.3in]{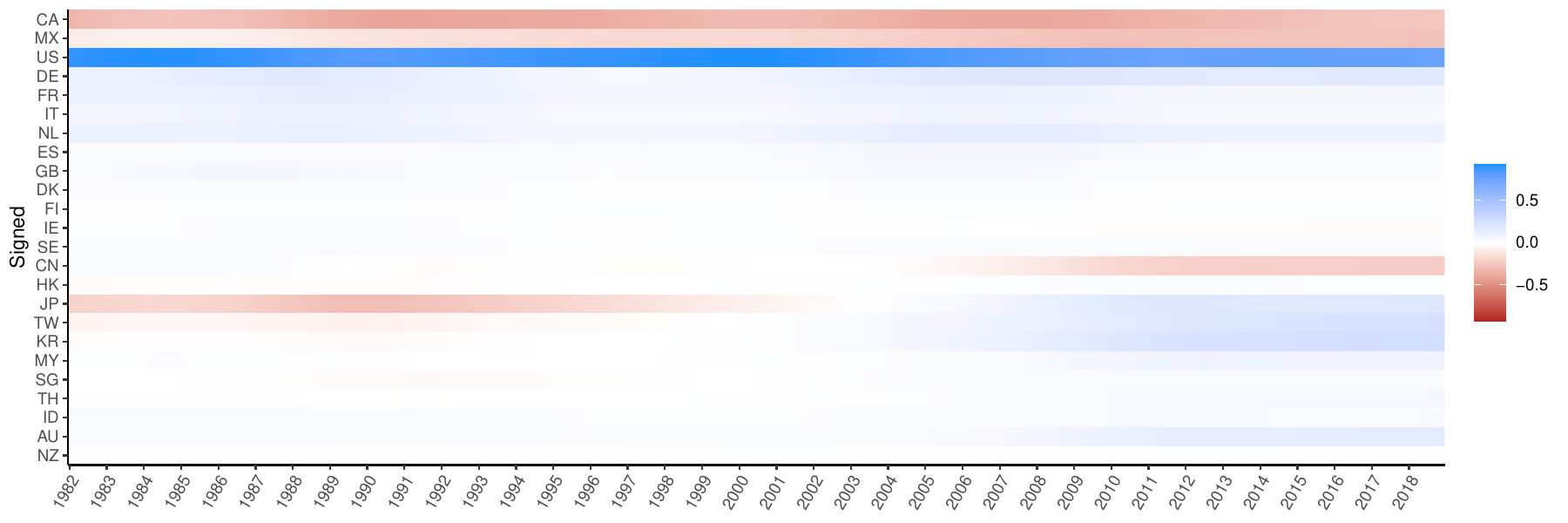}
\includegraphics[width=1\linewidth,height=1.3in]{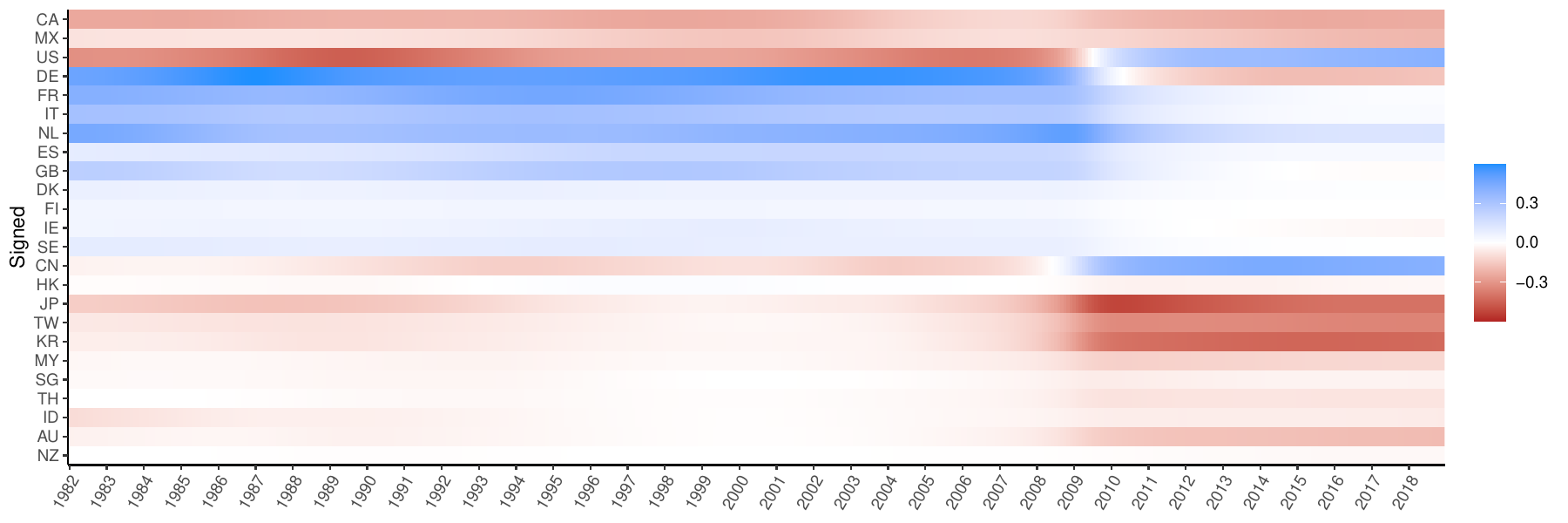}
\includegraphics[width=1\linewidth,height=1.3in]{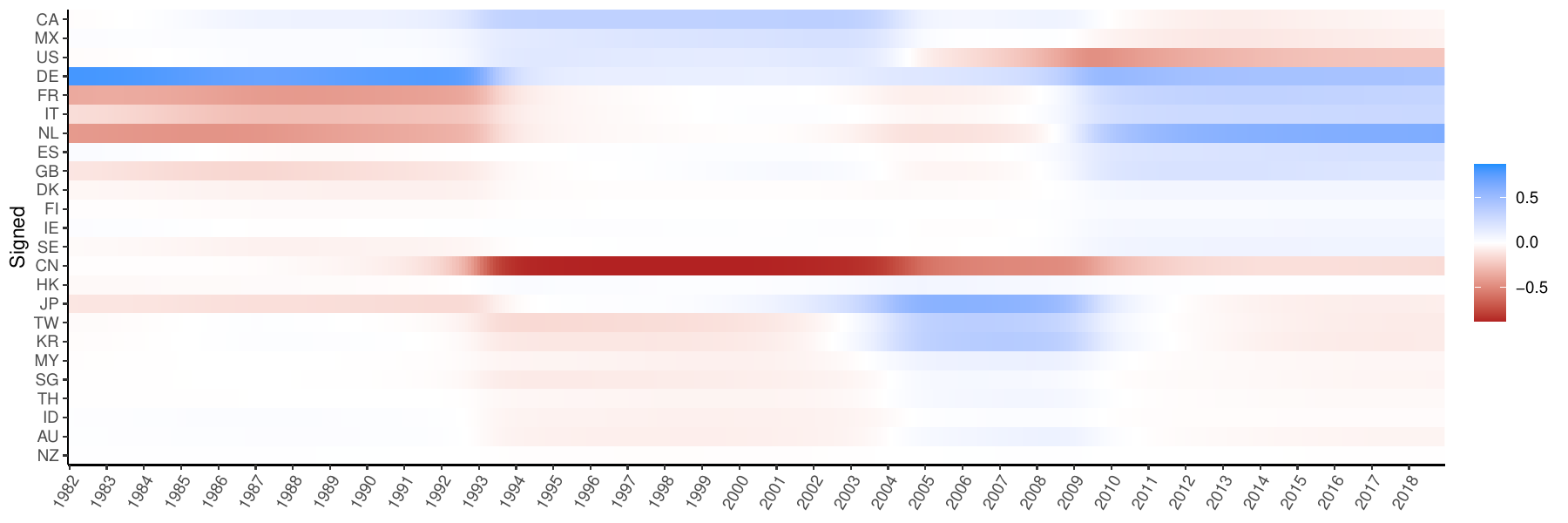}
\includegraphics[width=1\linewidth,height=1.3in]{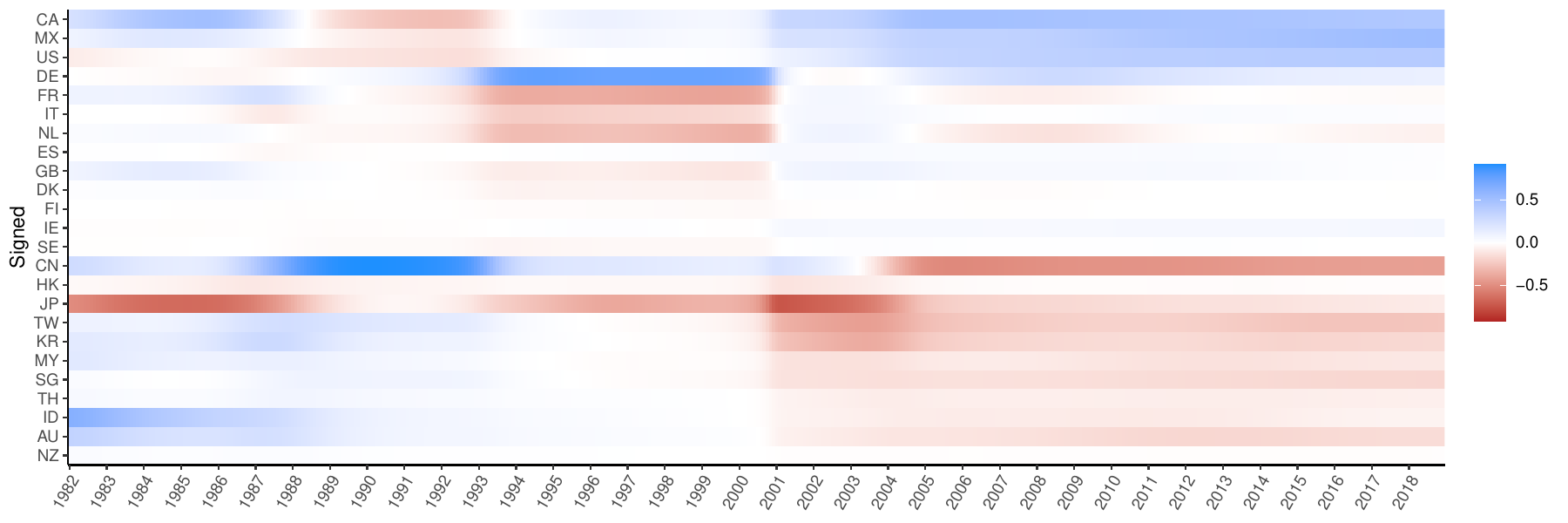}
\includegraphics[width=1\linewidth,height=1.3in]{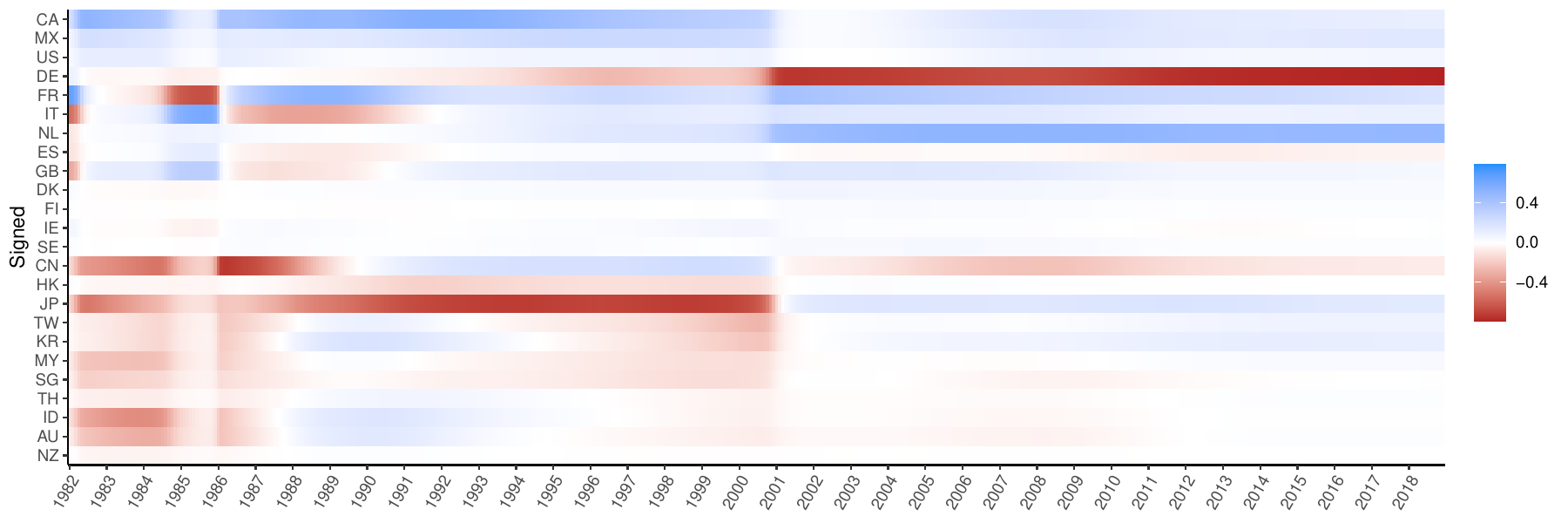}
\label{fig:export_loading_original}
\end{figure}

\begin{figure}[ht!]
\caption{Trading Volume Example: Test statistics $T_{i,t}$ of $\bR_t$. Horizontal line is the 0.95 upper IQR line.}
\includegraphics[width=1\linewidth,height=1.3in]{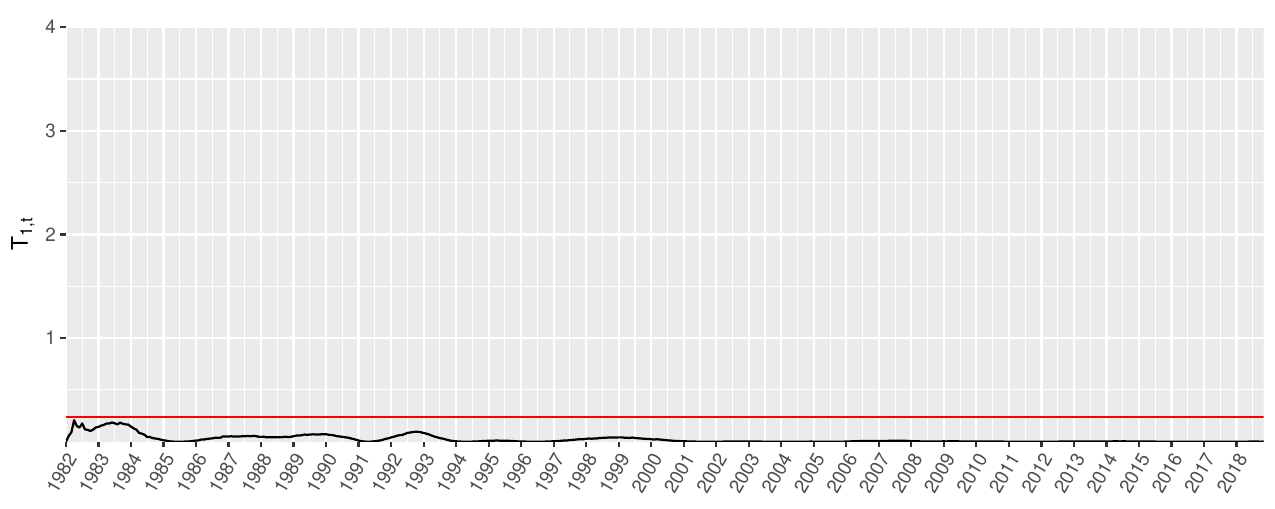}
\includegraphics[width=1\linewidth,height=1.3in]{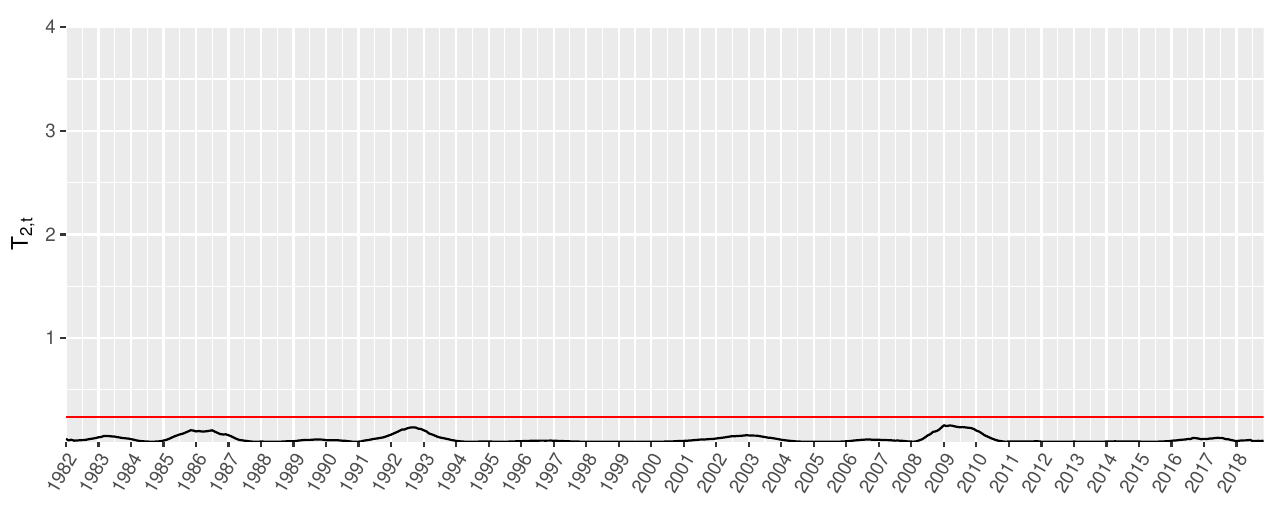}
\includegraphics[width=1\linewidth,height=1.3in]{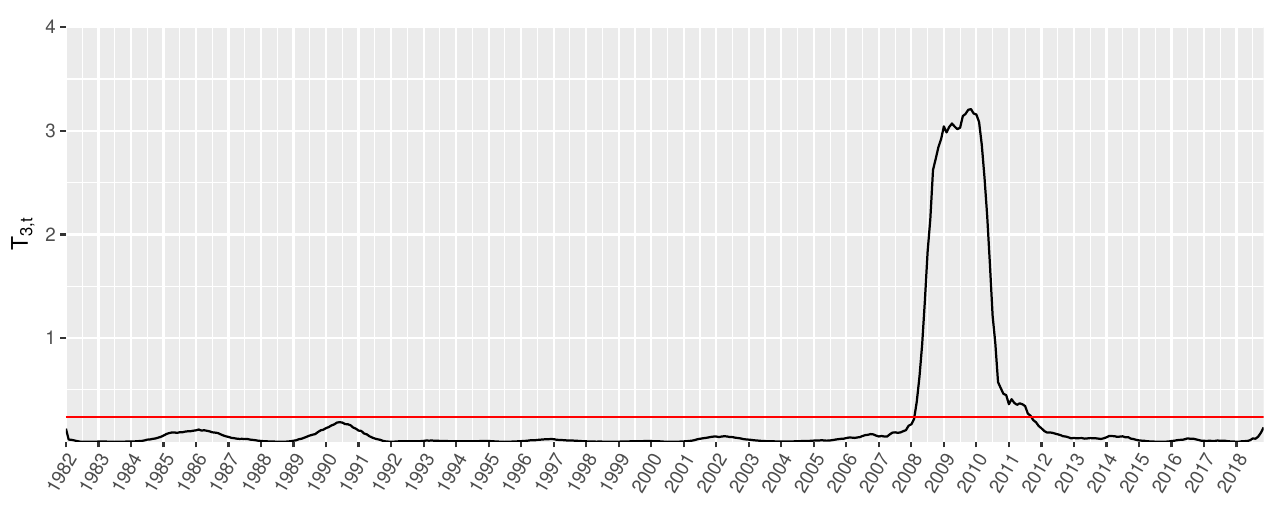}
\includegraphics[width=1\linewidth,height=1.3in]{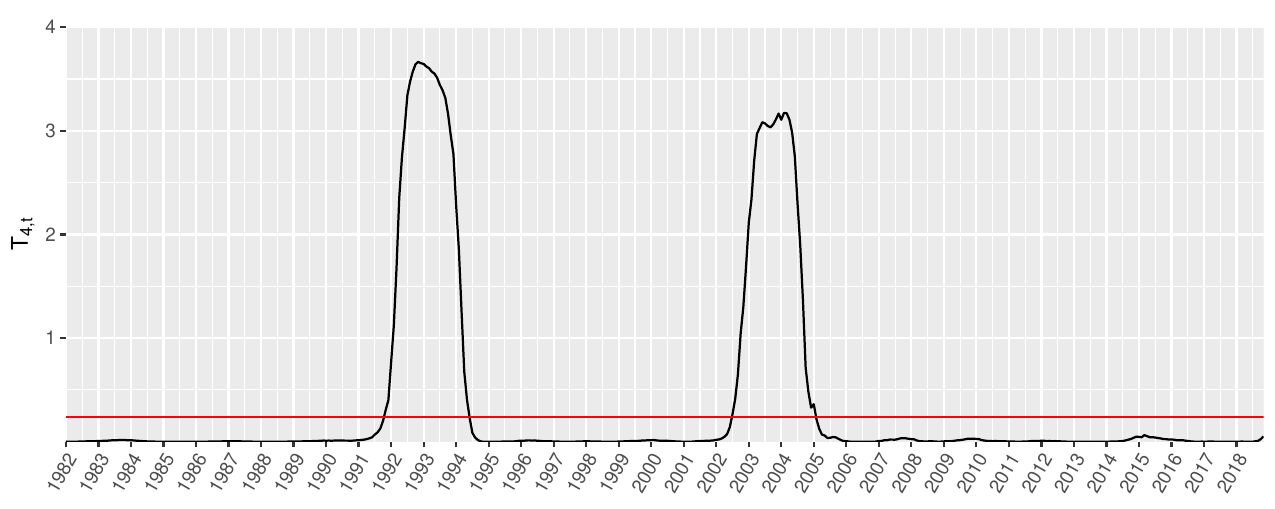}
\label{fig:export_TS}
\end{figure}

Figure \ref{fig:export_loading_original} shows the heatmap of the initial estimates of $\bR_t$, corresponding to the descending order of the eigenvalues. The first two eigenvectors (loading vectors) show quite smooth changes. However, the third and fourth eigenvectors show abrupt changes. Figure~\ref{fig:export_TS} shows the test statistics $T_{i,t}$ in \eqref{eq:test_i}, {where the large values indicate possible switching between $i$-th and $(i+1)$-th eigenvalues. 
It confirms no switching between the first two columns in $\bR_t$, but reveals switching between columns 3 and 4 around Year 2010, and two switches between columns 4 and 5 around Years 1993 and 2004.} Note that column 5 (corresponding to the fifth largest eigenvalue) was deemed insignificant 
by our model selection procedure. {It appears that the significance of the 5th eigenvalue of $\bM_{R,t}$ gradually increases, while the significance of the 4-th eigenvalue of $\bM_{R,t}$ gradually decreases to insignificance,} and they switch during the period. 

\begin{figure}[ht!]
\caption{Estimated eigenvalue functions}
\centerline{\includegraphics[width=0.8\linewidth,height=2.3in]{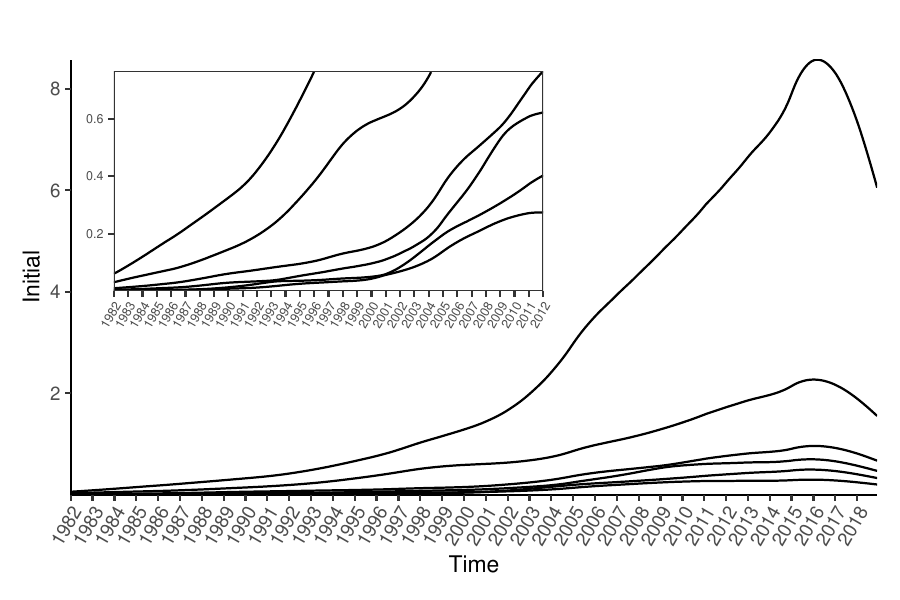}}
\label{fig:Eigenvalue_Rt1-1original}
\end{figure}

Due to the existence of switching, caution is required in labeling. Given the significant increase in import-export volume due to globalization, and thus increasing signal strengths, we adopt labels based on the later periods. That is, we name factor $k$ as the one associated with the largest $k$-th eigenvalue of $\bM_{R,T}$ where $T$ is the end of sample period. Figure~\ref{fig:Eigenvalue_Rt1-1original} shows the estimated six largest eigenvalues over time, suggesting potential switching around 2010 when the third and fourth eigenvalues are quite close. In the beginning when the signal is relatively weak, there are a number of potential switchings.

\begin{figure}[ht!]
\caption{Trading Volume Example: (top) Identified coalesing regions $\bR_t$ related to Factors with the original estimated after switching outside the coalescing region. (bottom) Result with interpolated loadings of Factor 3 within the coalescing region.}
\includegraphics[width=1\linewidth,height=1.3in]{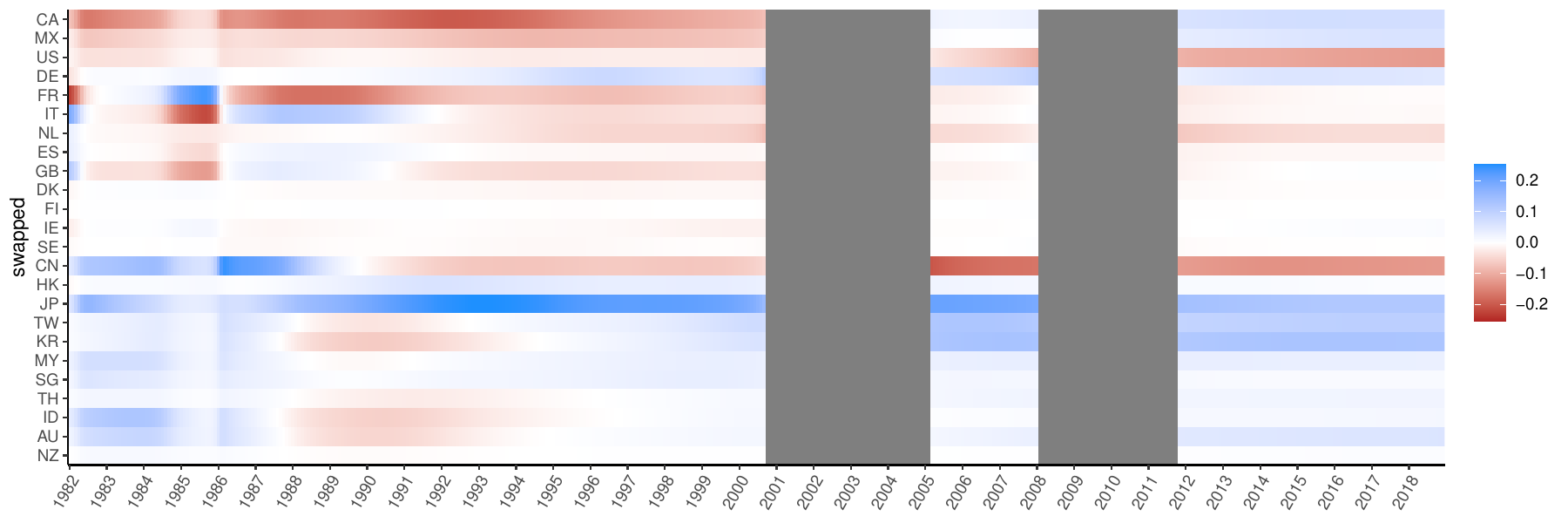}
\includegraphics[width=1\linewidth,height=1.3in]{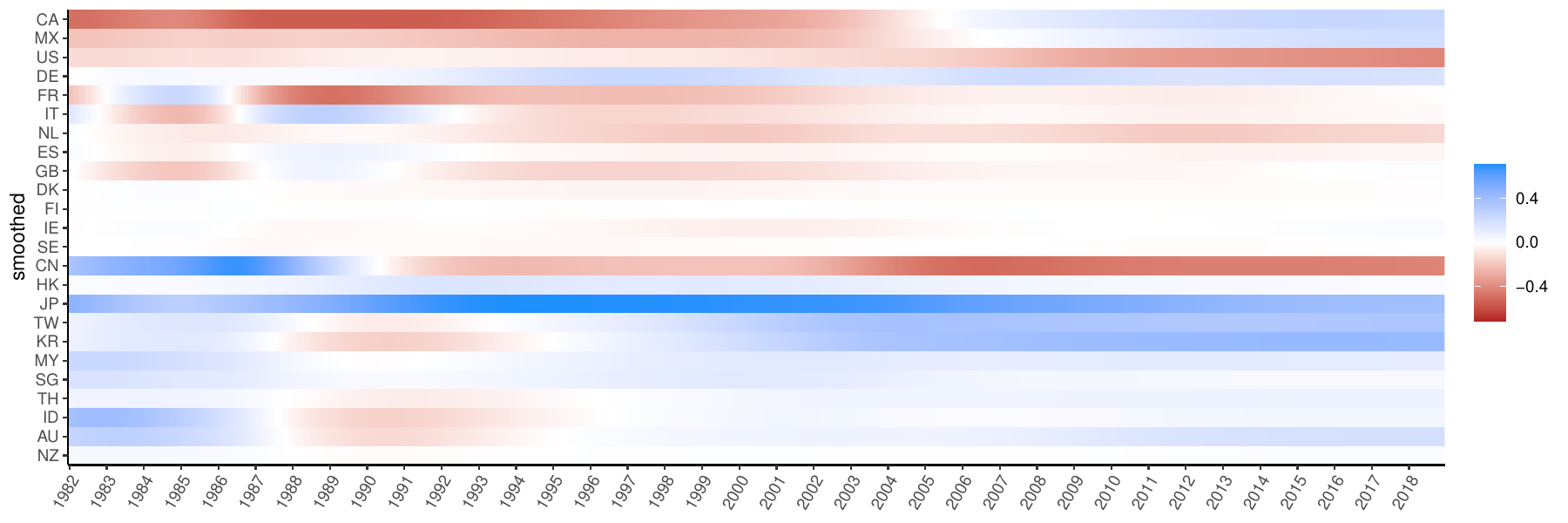}
\label{fig:export_swapping3}
\end{figure}

\begin{figure}[ht!]
\caption{Trading Volume Example: (top) Identified coalesing region $\bR_t$ between Factors 4 and 5, and switching outside the coalescing region. (bottom) Result with interpolated loadings within the coalescing region.}
\includegraphics[width=1\linewidth,height=1.3in]{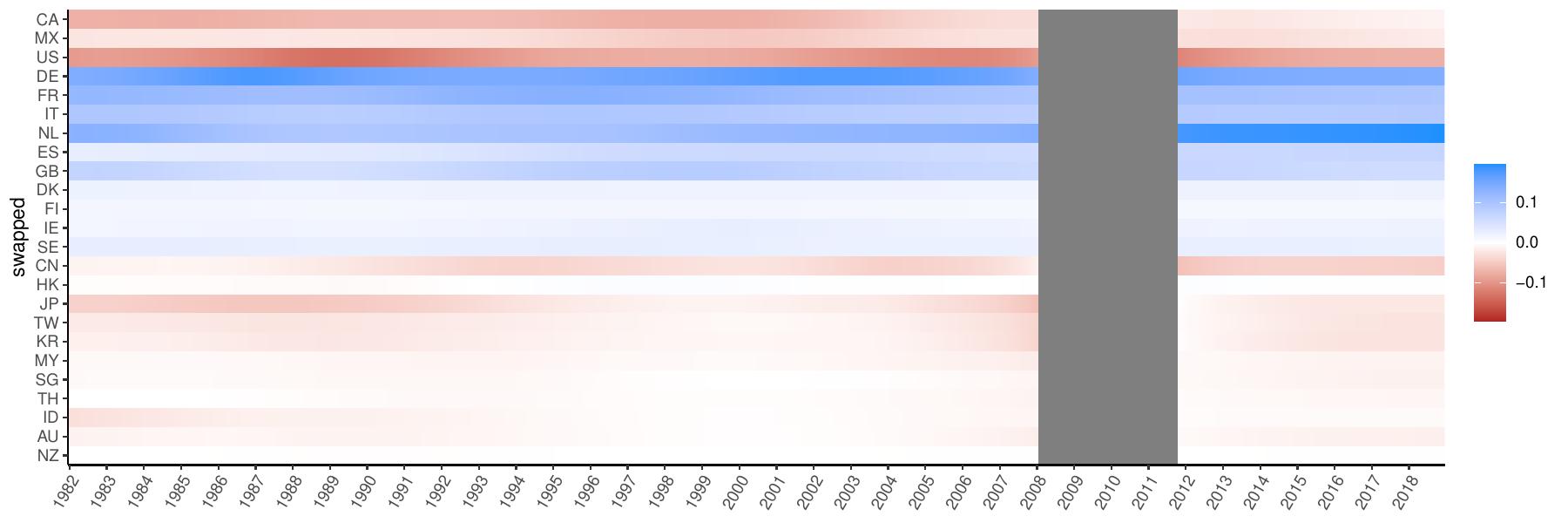}
\includegraphics[width=1\linewidth,height=1.3in]{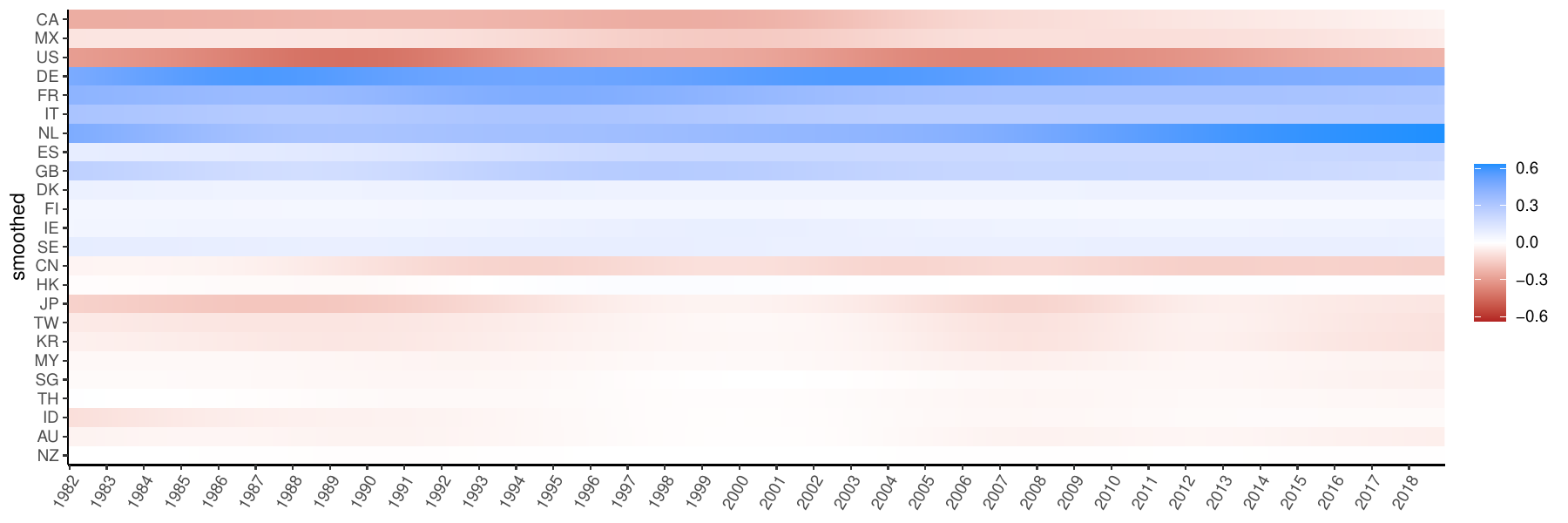}
\label{fig:export_swapping4}
\end{figure}

Using this labeling system, Figure~\ref{fig:export_TS} shows {a switch} between Factors 3 and 4 around Year 2010 (the third subfigure), implying that in the year {\bf after} 2010, Factor 3 corresponds to the 3rd largest eigenvalue of $\bM_{R,t}$ and its corresponding eigenvectors (as factor loading). Just before 2010, Factor 3 corresponds to the 4th largest eigenvalue, and it switches again around 2004 (the fourth subfigure), meaning that just before 2004, Factor 3 corresponds to the 5th largest eigenvalue.  The top figure in Figure~\ref{fig:export_swapping3} shows the identified coalescing region for Factors 3 in these two time periods, along with the initial estimates of the corresponding loading vectors {\bf after} the two switches outside the coalescing regions.  
The bottom figure shows the results after interpolating the loadings within the coalescing regions and smoothing outside the regions, following the procedure used in \cite{Motta2023}. Figure~\ref{fig:export_swapping4} shows the same for Factors 4, indicating a switch with Factor 3 around 2010. Both figures show that the smoothing operation in Section~\ref{sec:smooth} works well. 

\begin{figure}[ht!]
\caption{Trading volume example: Final estimates of $\bR_{t}$.}
\includegraphics[width=1\linewidth,height=1.3in]{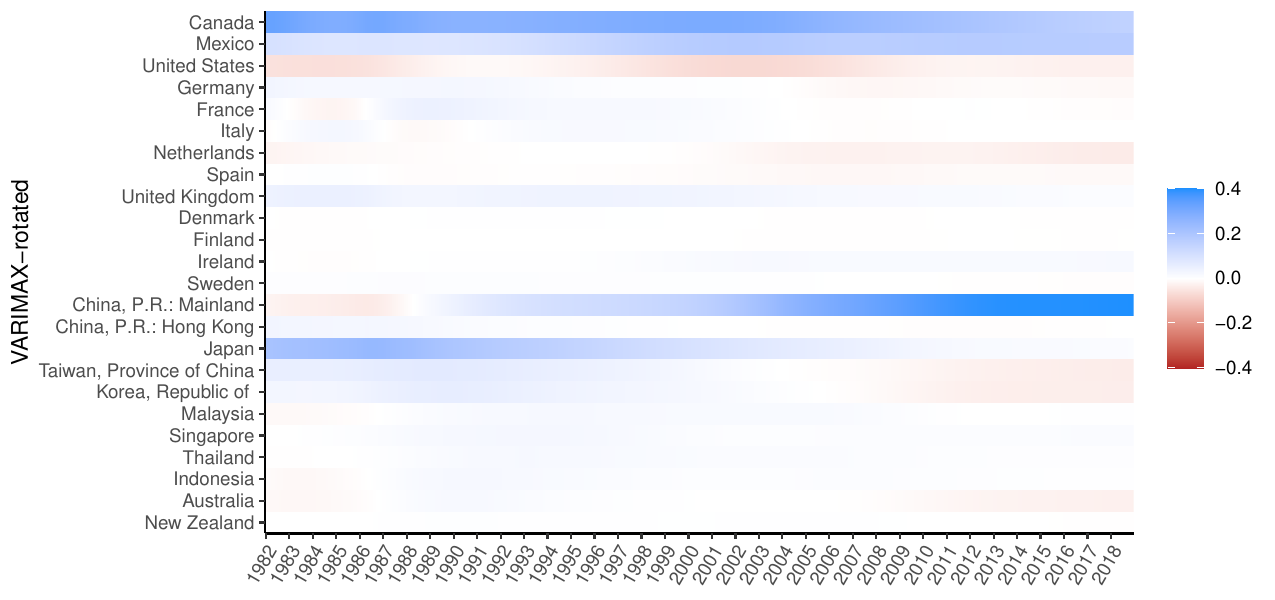}
\includegraphics[width=1\linewidth,height=1.3in]{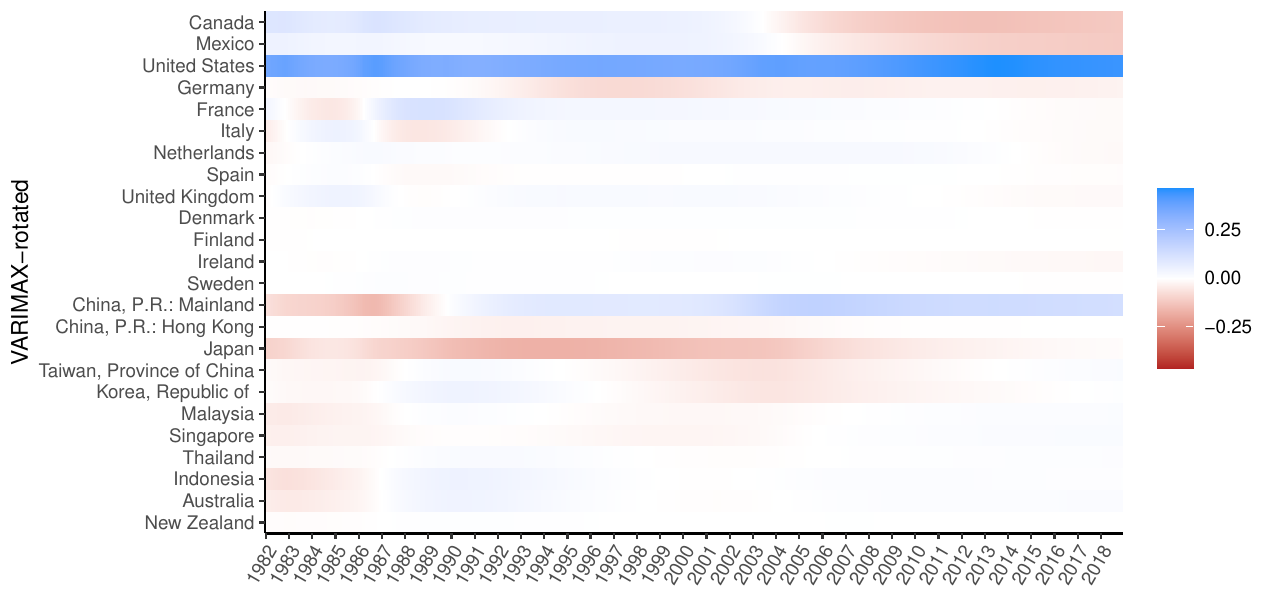}
\includegraphics[width=1\linewidth,height=1.3in]{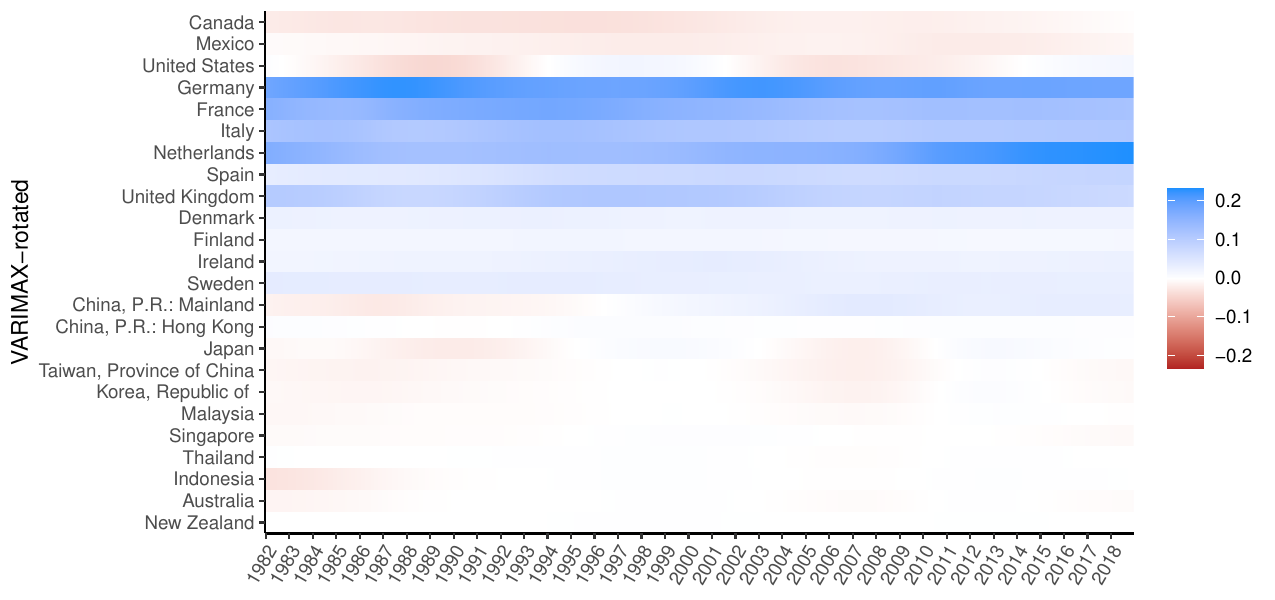}
\includegraphics[width=1\linewidth,height=1.3in]
{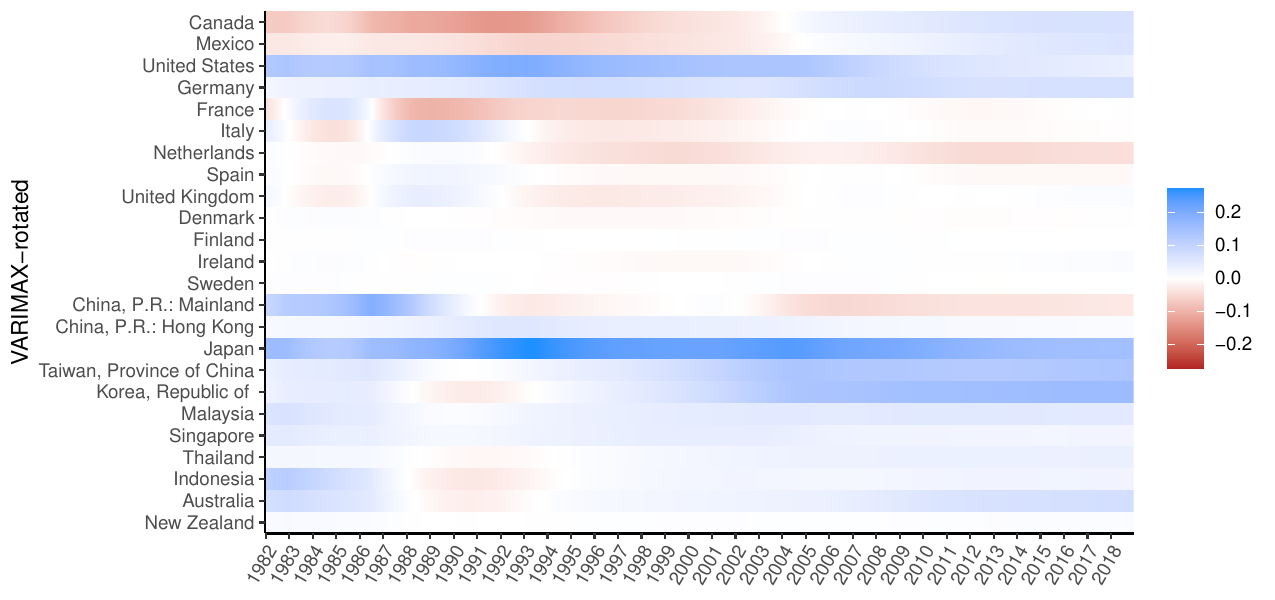}
\label{fig:export_loading_varimax}
\end{figure}

\begin{figure}[ht!]
\caption{Trading volume example: Final estimates of $\bC_{t}$.}
\includegraphics[width=1\linewidth,height=1.3in]{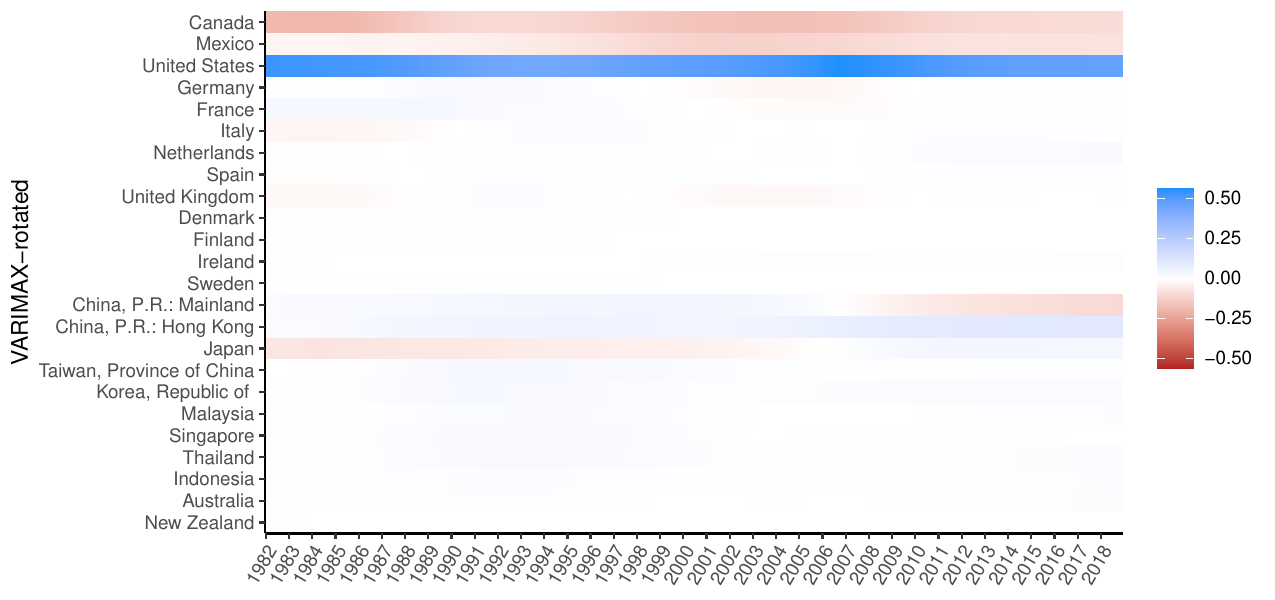}
\includegraphics[width=1\linewidth,height=1.3in]{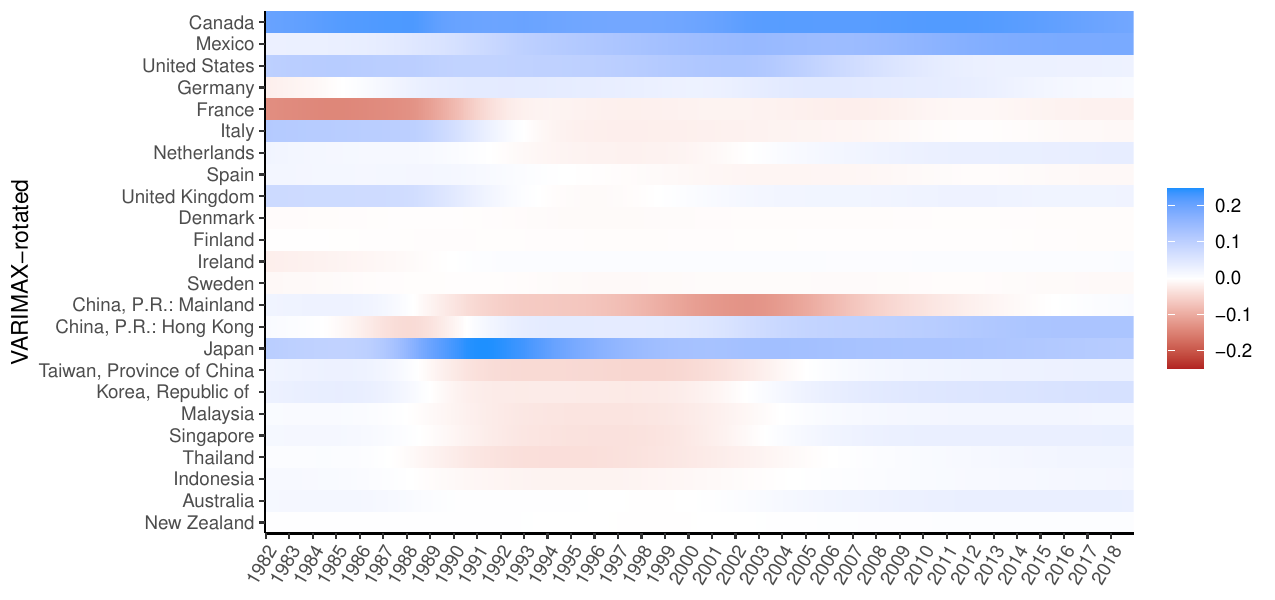}
\includegraphics[width=1\linewidth,height=1.3in]{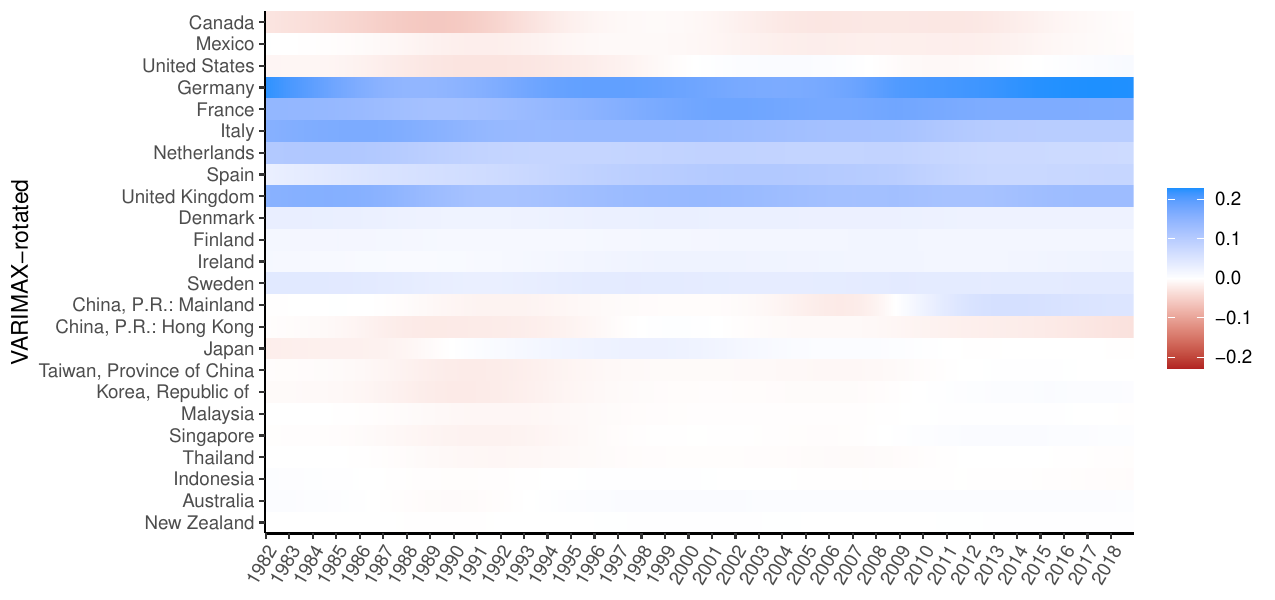}
\includegraphics[width=1\linewidth,height=1.3in]{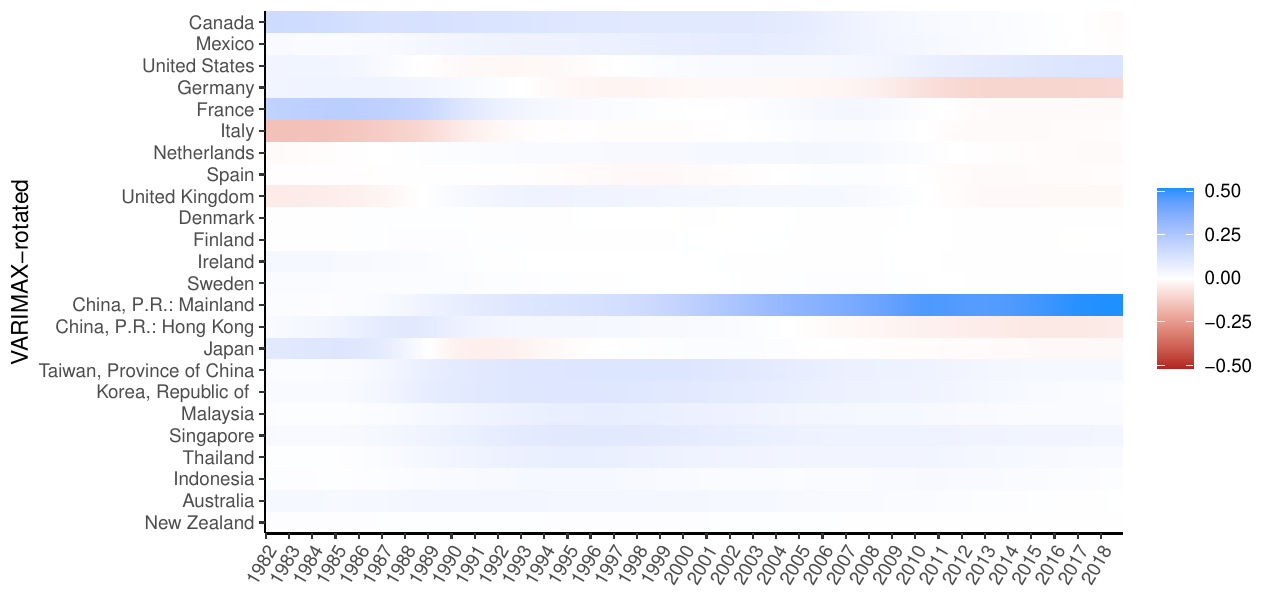}
\label{fig:import_loading_varimax}
\end{figure}

Since the signal is stronger in the later periods, we apply a global varimax rotation only using the smoothed eigenfunctions between 2003 to 2018. The rotation is then applied to the entire sample period.   
The final estimates of the four loading vector functions $\bR_t$ are shown in Figure~\ref{fig:export_loading_varimax}. Similar estimation approaches result in the final estimate of the loading vector functions of $\bC_t$, shown in Figure~\ref{fig:import_loading_varimax}. Note that after varimax rotation, the strength of the factor may not be in descending order anymore. 

\subsection{Interpretation of results}

Here we adopt the interpretation of the model described in Section~\ref{sec:model}. {Figures \ref{fig:export_loading_varimax} and \ref{fig:import_loading_varimax} provide insights into the dynamics of the globalization.}
The latent Export Hub 1 (top figure in Figure~\ref{fig:export_loading_varimax}) can be viewed as a China-dominated export hub, as China has the maximum loading overall. Before 2001, Japan had a higher loading on this hub than China, indicating a shift in exporting centrality within the Asian economy. China's loading on this hub experiences significant growth, particularly following its accession to the WTO in 2001. Despite this, Japan, Canada, Hong Kong and Taiwan also remain significant contributors to this export hub.

The latent Export Hub 2 can be viewed as a USA-dominated export hub as {evidenced by the substantially larger loading of the USA compared to other countries. While China also participates in this hub, the United States maintains dominance throughout the sample period.
The latent Export Hub 3 can be interpreted as a hub dominated by European countries such as Germany, France, Italy, Netherlands, and Spain. Last, the latent Export Hub 4 was mainly used by APEC countries, in which Japan dominated throughout the period, {while Korea, Taiwan and Hong Kong emerge as active participants post-2000.} United States also participated in this hub, {albeit to a lesser extent}.

On the import side, United States dominated Import Hub 1 (top figure in Figure~\ref{fig:import_loading_varimax}). Canada, Japan and Mexico used Import Hub 2. The European countries used Import Hub 3, and China dominated in Import Hub 4.

\section{Conclusion} \label{sec:conclusion}

Modeling high-dimensional matrix-valued time series has attracted increasing attention recently. We contribute to this literature by proposing a new time-varying matrix factor model that may be more applicable in real applications, establishing the asymptotic properties of a local PCA estimator for the model, and proposing a generalized eigenvalue-ratio estimator for latent dimensions. Our proposed estimators are intuitively appealing and straightforward to compute. They fully preserve the matrix structure and explore the local variation of the loading matrices. Moreover, our results are obtained under very general conditions that allow for weak correlations across time, rows, and columns. {A smoothing operation is used to obtain a smooth representation of the time-varying loading matrices in order to facilitate easy interpretation of the model. Simulations have shown that the nonparametric estimator for factor loadings and the generalized eigenvalue ratio-based estimator perform well in finite samples. The empirical application concerning the international trade flow reveals some interesting time-varying patterns.

\spacingset{1}
\bibliographystyle{agsm}
\bibliography{main}

\end{document}